\documentclass[sigconf]{acmart}
\setcopyright{acmcopyright}
\usepackage{booktabs} % For formal tables
\usepackage{microtype}
\usepackage[latin1]{inputenc}
\usepackage[OT2,T1]{fontenc}
\usepackage{amssymb}
\usepackage[linesnumbered,ruled,vlined]{algorithm2e}
\usepackage{algpseudocode}
\usepackage{bm}
\usepackage{amsmath}
\usepackage{graphicx}
\usepackage{epsfig}
\usepackage{url}
\usepackage{psfrag}
\usepackage{balance}
\usepackage{float}
\usepackage{multirow}
\usepackage{pstricks}
\usepackage{array}
\usepackage{enumitem}
\usepackage{hyperref}
\usepackage{pifont}
\usepackage{bbm}
\usepackage{caption}
\usepackage{subcaption}
\graphicspath{{figs/}}

\def\BibTeX{{\rm B\kern-.05em{\sc i\kern-.025em b}\kern-.08emT\kern-.1667em\lower.7ex\hbox{E}\kern-.125emX}}

\copyrightyear{2020}
\acmYear{2020}
\setcopyright{iw3c2w3}
\acmConference[WWW '20]{Proceedings of The Web Conference 2020}{April 20--24, 2020}{Taipei, Taiwan}
\acmBooktitle{Proceedings of The Web Conference 2020 (WWW '20), April 20--24, 2020, Taipei, Taiwan}
\acmPrice{}
\acmDOI{10.1145/3366423.3380160}
\acmISBN{978-1-4503-7023-3/20/04}

\begin{document}

\title{Fast Generating A Large Number of Gumbel-Max Variables}

\settopmatter{printccs=true, printacmref=false, printfolios=false}
\author{
	Yiyan Qi$^{1}$, Pinghui Wang$^{2,1,*}$, Yuanming Zhang$^{1}$, Junzhou Zhao$^{1,*}$,  
Guangjian Tian$^{3}$,}
\author{Xiaohong Guan$^{2,1,4}$}
\thanks{$^*$Corresponding Author.}
\affiliation{%
	\institution{$^{1}$MOE Key Laboratory for Intelligent Networks
		and Network Security, Xi'an Jiaotong University, Xi'an, China}
	\institution{$^{2}$Shenzhen Research Institute of Xi'an Jiaotong University, Shenzhen, China}
	\institution{$^{3}$Huawei Noah's Ark Lab, Hong Kong}
	\institution{$^{4}$Department of Automation and NLIST Lab, Tsinghua University, Beijing, China}
}
\email{{qiyiyan,zhangyuanming}@stu.xjtu.edu.cn, {phwang,xhguan}@mail.xjtu.edu.cn,} \email{junzhou.zhao@xjtu.edu.cn, Tian.Guangjian@huawei.com}

\renewcommand{\shortauthors}{Yiyan Qi, et al.}

\begin{abstract}
The well-known Gumbel-Max Trick for sampling elements from a categorical distribution (or more generally a nonnegative vector) and its variants have
been widely used in areas such as machine learning and information retrieval.
To sample a random element $i$ (or a Gumbel-Max variable $i$) in proportion to its positive weight $v_i$,
the Gumbel-Max Trick first computes a Gumbel random variable $g_i$ for each positive weight element $i$, and then samples the element $i$ with the largest value of $g_i+\ln v_i$.
Recently,
applications including similarity estimation and graph embedding require
to generate $k$ independent Gumbel-Max variables from high dimensional vectors.
However, it is computationally expensive for a large $k$ (e.g., hundreds or even thousands) when using the traditional Gumbel-Max Trick.
To solve this problem, we propose a novel algorithm, \emph{FastGM}, that reduces
the time complexity from $O(kn^+)$ to $O(k \ln k + n^+)$,
where $n^+$ is the number of positive elements in the vector of interest.
Instead of computing $k$ independent Gumbel random variables directly,
we find that there exists a technique to generate these variables in descending order.
Using this technique, our method FastGM computes variables $g_i+\ln v_i$ for all positive elements $i$ in descending order.
As a result, FastGM significantly reduces the computation time because we can stop the procedure of Gumbel random variables computing for many elements especially for those with small weights.
Experiments on a variety of real-world datasets show that FastGM is orders of magnitude faster than state-of-the-art methods without sacrificing accuracy and incurring additional expenses.
\end{abstract}



%
% Keywords. The author(s) should pick words that accurately describe the work being
% presented. Separate the keywords with commas.

\maketitle
{\fontsize{8pt}{8pt} \selectfont
	\textbf{ACM Reference Format:}\\
	Yiyan Qi, Pinghui Wang, Yuanming Zhang, Junzhou Zhao, Guangjian Tian, and Xiaohong Guan. 2020.Fast Algorithm for Generating A Large Number of Gumbel-Max Variables. In \text{\it Proceedings of The Web Conference
		2020}
	\text{\it(WWW '20), April 20-24, 2020, Taipei, Taiwan.} ACM, New York, NY, USA, 11 pages. https://doi.org/10.1145/3366423.3380160}

\section{Introduction} \label{sec:introduction}

The Gumbel-Max Trick~\cite{luce2012individual} is a popular technique for sampling elements
from a categorical distribution (or more generally a nonnegative vector).
Given a nonnegative vector $\vec{v}=(v_1,\ldots,v_n)$ where each element
$v_i\in\mathbb{R}_{\geq 0}$, the Gumbel-Max Trick computes a random variable
$s(\vec{v})$ as
\begin{equation*}
  s(\vec{v}) = \arg\max_{i\in N^+_{\vec{v}}} g_i + \ln v_i,
\end{equation*}
where $N^+_{\vec{v}}\triangleq\{i\colon v_i > 0, i=1,\ldots,n\}$ is the set of
indices of positive elements in $\vec{v}$, $g_i\triangleq -\ln(-\ln a_i)$ and
$a_i$ is a random variable drawn from the uniform distribution $\text{UNI}(0,1)$.
We call $s(\vec{v})$ a Gumbel-Max variable of vector $\vec{v}$ and the probability
of selecting $i$ as the Gumbel-Max variable is $P(s(\vec{v}) = i) =
\frac{v_i}{\sum_{j=1}^n v_j}$.
The Gumbel-Max Trick and its variants have been used widely in many areas.

\textbf{Similarity estimation.}
Similarity estimation lies at the core of many data mining and machine learning
applications, such as web duplicate detection
\cite{henzinger2006finding,manku2007detecting}, collaborate filtering
\cite{bachrach2009sketching}, and association rule learning
\cite{MitzenmacherWWW14}.
To efficiently estimate the similarity between two vectors, several
algorithms~\cite{yang2016poisketch,yang2017histosketch,yang2018d2,moulton2018maximally}
compute $k$ random variables $-\frac{\ln a_{i,1}}{v_i}, \ldots, -\frac{\ln
	a_{i,k}}{v_i}$ for each positive element $v_i$ in $\vec{v}$, where $a_{i,1},
\ldots, a_{i,k}$ are independent random variables drawn from the uniform
distribution $\text{UNI}(0,1)$.
Then, these algorithms build a sketch (or called \emph{Gumbel-Max sketch} in this
paper) of vector $\vec{v}$ consisting of $k$ registers, and each register records
$s_j(\vec{v})$ where
\[
s_j(\vec{v}) = \arg\min_{i\in N^+_{\vec{v}}} - \frac{\ln a_{i,j}}{v_i},
\quad 1 \le j \le k.
\]
We find that $s_j(\vec{v})$ is exactly a Gumbel-Max variable of vector $\vec{v}$
as $\arg\min_{i\in N^+_{\vec{v}}} - \frac{\ln a_{i,j}}{v_i} = \arg\max_{i\in
	N^+_{\vec{v}}} \ln v_i - \ln (- \ln a_{i, j})$.
Let $\mathbbm{1}(x)$ be an indicator function.
Yang et al.~\cite{yang2016poisketch,yang2017histosketch,yang2018d2} use
$\frac{1}{k}\sum_{j=1}^k \mathbbm{1}(s_j(\vec{u}) = s_j(\vec{v}))$ to estimate the
\emph{weighted Jaccard similarity} of two nonnegative vectors $\vec{u}$ and
$\vec{v}$ which is defined by
\[
\mathcal{J_W}(\vec{u},\vec{v})\triangleq
\frac{\sum_{i=1}^n \min\{u_i, v_i\}}{\sum_{i=1}^n \max\{u_i, v_i\}}.
\]
Recently, Moulton et al.~\cite{moulton2018maximally} prove that the expectation of
estimate $\frac{1}{k}\sum_1^k \mathbbm{1}(s_j(\vec{u}) = s_j(\vec{v}))$ actually
equals the \emph{probability Jaccard similarity}, which is defined by
\[
\mathcal{J_P}(\vec{u}, \vec{v})\triangleq
\sum_{i\in N^+_{\vec{v},\vec{u}}}
\frac{1}{\sum_{l=1}^n \max\left(\frac{u_l}{u_i}, \frac{v_l}{v_i}\right)}.
\]
Here, $N^+_{\vec{v},\vec{u}}\triangleq\{i\colon v_i > 0\wedge u_i > 0, i=1,
\ldots, n\}$ is the set of indices of positive elements in both $\vec{v}$ and
$\vec{u}$.
Compared with the weighted Jaccard similarity $\mathcal{J_W}$, Moulton et al. demonstrate that the probability Jaccard similarity $\mathcal{J_P}$ is
scale-invariant and more sensitive to changes in vectors.
Moreover, each function $s_j(\vec{v})$ maps similar vectors into the same value
with high probability.
Therefore, similar to the regular locality-sensitive hashing (LSH)
schemes~\cite{GionisPVLDB1999,Broder2000,Charikar2002similarity}, one can use
these Gumbel-Max sketches to build LSH index for fast similarity search in a large
dataset, which is capable to search similar vectors for any query vector in
sub-linear time.

\textbf{Graph embedding.}
Recently, graph embedding attracts a lot of attention.
A variety of graph embedding methods have been developed to transform a graph into
a low-dimensional space in which each node is represented by a low-dimensional
vector and meanwhile the proximities of nodes are preserved.
With node embeddings, many off-the-shelf data mining and machine learning
algorithms can be applied on graphs such as node classification and link
prediction.
Lately, Yang et al.~\cite{yang2019nodesketch} reveal that existing methods are
computationally intensive especially for large graphs.
To address this challenge, Yang et al.
build a Gumbel-Max sketch described above for each row of the graph's
self-loop-augmented adjacency matrix and use these sketches as the first-order
node embeddings of nodes.
To capture the high-order node proximity, they propose a fast method
\emph{NodeSketch} to recursively generate $r$-order node embeddings (i.e.,
Gumbel-Max sketches) based on the graph's self-loop-augmented adjacency matrix and
$(r-1)$-order node embeddings.

\textbf{Machine learning.}
Lorberbom et al.~\cite{lorberbom2018direct} use the Gumbel-Max Trick to
reparameterize discrete variant auto-encoding (VAE).
Buchnik et al.~\cite{buchnik2019self} apply the Gumbel-Max Trick to select
training examples, and sub-epochs of sampled examples preserve rich structural
properties of the full training data.
These studies show that the trick not only significantly accelerates the
convergence of the loss function, but also improves the accuracy.
Other examples include reinforcement learning~\cite{oberst2019counterfactual}, and
integer linear programs~\cite{kim2016exact}.
Lately, Eric et al.~\cite{jang2016categorical} extend the Gumbel-Max Trick to
embed discrete Gumbel-Max variables in a continuous space, which enables us to
compute the gradients of these random variables easily.
This technique has also been used for improving the performance of neural network
models such as Generative Adversarial Networks (GAN)~\cite{kusner2016gans} and attention models~\cite{tay2018multi}.

Despite the wide use of the Gumbel-Max Trick in various domains, it is expensive
for the above algorithms to handle large dimensional vectors.
Specifically, the time complexity of generating a Gumbel-Max sketch consisting of
$k$ independent Gumbel-Max variables is $O(n^+k)$, where $n^+=|N^+|$ is the number
of positive elements in the vector of interest.
In practice, $k$ is usually set to be hundreds or even thousands when selecting
training samples \cite{buchnik2019self}, estimating probability Jaccard similarity
\cite{moulton2018maximally}, and learning NodeSketch graph embedding
\cite{yang2019nodesketch}.
To solve this problem, we propose a novel method, \emph{FastGM}, to fast compute a
Gumbel-Max sketch.
The basic idea behind FastGM can be summarized as follows.
For each element $v_i>0$ in $\vec{v}$, we find that $k$ random variables
$-\frac{\ln a_{i,1}}{v_i}, \ldots, - \frac{\ln a_{i,k}}{v_i}$ can be generated in
ascending order.
That is, we can generate a sequence of $k$ tuples $\left(-\frac{\ln a_{i, i_1}}
  {v_i}, i_1\right), \ldots, \left(- \frac{\ln a_{i, i_k}}{v_i}, i_k\right)$,
where $-\frac{\ln a_{i, i_1}} {v_i}< \cdots < -\frac{\ln a_{i, i_k}} {v_i}$ and
$i_1, \ldots, i_k$ is a random permutation of integers $1,\ldots, k$.
We propose to sort $kn^+$ random variables of all $n^+$ positive elements and
compute these variables sequentially in ascending order.
Then, we model the procedure of computing the Gumbel-Max sketch $(s_1(\vec{v}),
\ldots, s_k(\vec{v}))$ of vector $\vec{v}$ as a Balls-and-Bins model.
Specifically, randomly throw balls one by one into $k$ empty bins, where
each ball is assigned a random variable in ascending order.
When no bins are empty, we early stop the procedure, and then each $s_j(\vec{v})$,
$j=1, \ldots, k$, records the random variable of the first ball thrown into bin
$j$.
We summarize our main contributions as:
\begin{itemize}[leftmargin=*]
\item We introduce a simple Balls-and-Bins model to interpret the procedure of
  computing the Gumbel-Max sketch of vector $\vec{v}$, i.e., $(s_1(\vec{v}),
  \ldots, s_k(\vec{v}))$.
  Using this stochastic process model, we propose a novel algorithm, called
  FastGM, to reduce the time complexity of computing the Gumbel-Max sketch
  $(s_1(\vec{v}), \ldots, s_k(\vec{v}))$ from $O(n^+k)$ to $O(k \ln k + n^+)$,
  which is achieved by avoiding calculating all $k$ variables $- \frac{\ln
    a_{i,1}}{v_i}, \ldots, - \frac{\ln a_{i,k}}{v_i}$ for each $i \in
  N^+_{\vec{v}}$.
%  (Sections~\ref{sec:problem} and~\ref{sec:method})
\item We conduct experiments on a variety of real-world datasets for estimating
  probability Jaccard similarity and learning NodeSketch graph embeddings.
  The experimental results demonstrate that our method FastGM is orders of
  magnitude faster than the state-of-the-art methods without incurring any
  additional cost.
%  (Section~\ref{sec:results})
\end{itemize}

The rest of this paper is organized as follows.
Section~\ref{sec:related} summarizes related work.
The problem formulation is presented in Section~\ref{sec:problem}.
Section~\ref{sec:method} presents our method FastGM.
The performance evaluation and testing results are presented in
Section~\ref{sec:results}.
Concluding remarks then follow.

%%% Local Variables:
%%% mode: latex
%%% TeX-master: "FastGM"
%%% End:

\section{Related Work} \label{sec:related}

%{\color{red}
%\subsection{Gumbel-Max Samples}
%}

\subsection{Jaccard Similarity Estimation}

% To estimate the Jaccard similarity between two sets (or binary vectors),
Broder et al.~\cite{Broder2000} proposed the first sketch method \emph{MinHash} to
compute the Jaccard similarity of two sets (or binary vectors).
MinHash builds a sketch consisting of $k$ registers for each set.
Each register uses a hash function to keep track of the set's element with the minimal hash
value.
To further improve the performance of MinHash,
\cite{PingWWW2010,MitzenmacherWWW14,Wang2019mem} developed several
memory-efficient methods.
Li et al.~\cite{Linips2012} proposed \emph{One Permutation Hash} (OPH) to reduce the time
complexity of processing each element from $O(k)$ to $O(1)$ but this method may exhibit large estimation errors because of the empty buckets.
To solve this problem, several densification
methods~\cite{ShrivastavaUAI2014,ShrivastavaICML2014,ShrivastavaICML2017,dahlgaard2017fast}
were developed to set the registers of empty buckets according to the values of
non-empty buckets' registers.

Besides binary vectors, a variety of methods have also been developed to estimate
generalized Jaccard similarity on weighted vectors.
For vectors consisting of only nonnegative integer weights, Haveliwala et
al.~\cite{haveliwala2000scalable} proposed to add a corresponding number of
replications of each element in order to apply the conventional MinHash.
To handle more general real weights, Haeupler et al.~\cite{haeupler2014consistent}
proposed to generate another additional replication with probability that equals
the floating part of an element's weight.
These two algorithms are computationally intensive when computing hash values of
massive replications for elements with large weights.
To solve this problem, \cite{gollapudi2006exploiting, Manasse2010} proposed to
compute hash values only for few necessary replications (i.e., ``active
indices'').
ICWS~\cite{IoffeICDM2010} and its variations such as 0-bit CWS~\cite{LiKDD2015},
CCWS~\cite{WuICDM2016}, PCWS~\cite{WuWWW2017}, I$^2$CWS~\cite{wu2018improved} were
proposed to improve the performance of CWS~\cite{Manasse2010}.
% In addition, Chum et.
% al.~\cite{chum2008near} employed exponential distribution to make each element
% sampled proportionally to its weight.
% However, this algorithm also offers a biased estimate to the weighted Jaccard
% similarity.
% Given weight upper bounds as a priori knowledge, Shrivastava
% \cite{Shrivastava2016simple} proposed a simple weighted MinHash algorithm by
% using rejection sampling.
The CWS algorithm and its variants all have the time complexity of $O(n^+k)$,
where $n^+$ is the number of elements with positive weights.
Recently, Otmar~\cite{ertl2018bagminhash} proposed another efficient algorithm
\emph{BagMinHash} for handling high dimensional vectors.
BagMinHash is faster than ICWS when the vector has a large number of positive
elements, e.g., $n^+>1,000$, which may not hold for many real-world datasets.
The above methods all estimate the weighted Jaccard similarity.
Ryan et al.~\cite{moulton2018maximally} proposed a Gumbel-Max Trick based
sketching method, $\mathcal{P}$-MinHash, to estimate another novel Jaccard
similarity metric, \emph{probability Jaccard similarity}.
They also demonstrated that the probability Jaccard similarity is scale-invariant
and more sensitive to changes in vectors.
However, the time complexity of $\mathcal{P}$-MinHash processing a weighted vector
is $O(n^+k)$, which is not feasible for high-dimensional vectors.

\subsection{Graph Embedding}
DeepWalk \cite{perozzi2014deepwalk} employed truncated random-walks to transform a
network into sequences of nodes and learns node embedding using skip-gram model
\cite{Mikolov2013Distributed}.
The basic idea behind DeepWalk is that two nodes should have similar embedding
when they tend to co-occur in short random-walks.
Node2vec~\cite{Grover2016node2vec} extended DeepWalk by introducing two kinds of
search strategies, i.e., breadth- and depth-first search, into random-walk.
LINE~\cite{tang2015line} explicitly defined the first-order and second-order
proximities between nodes, and learned node embedding by minimizing the two
proximities.
GraRep~\cite{Cao2015GraRep} extended LINE to high-order proximities.
Qiu et al.~\cite{qiu2017network} unified the above methods into a general matrix
factorization framework.
Many recent works also learned embeddings on attributed graphs \cite{li2017attributed,dave2019neural,bonner2019exploring,lan2020improving}, partially labeled graphs \cite{tu2016max, li2016discriminative,yang2016revisiting,kipf2016semi}, and dynamic graphs \cite{Nguyen2018Continuous,Yu2018netwalk,du2018dynamic,qi2019discriminative}.
The above methods are computationally intensive and so many are prohibitive for large graphs.
To solve this problem, recently, Yang et al.~\cite{yang2019nodesketch} developed a
novel embedding method, NodeSketch, which builds a Gumbel-Max sketch for each row
$i$ of the graph's self-loop-augmented adjacency matrix and uses it as the
low-order node embedding of node $i$.
To capture the high-order node proximity, NodeSketch recursively generates
$r$-order node embeddings (i.e., Gumbel-Max sketches) based on the graph's
self-loop-augmented adjacency matrix and $(r-1)$-order node embeddings.
NodeSketch requires time complexity $O(n^+k)$ to compute the Gumbel-Max sketch of
a node, which is expensive when existing a large number of nodes in the graph.

%%% Local Variables:
%%% mode: latex
%%% TeX-master: "FastGM"
%%% End:

\section{Problem Formulation} \label{sec:problem}

\begin{figure*}[t]
	\center
	\includegraphics[width=\textwidth]{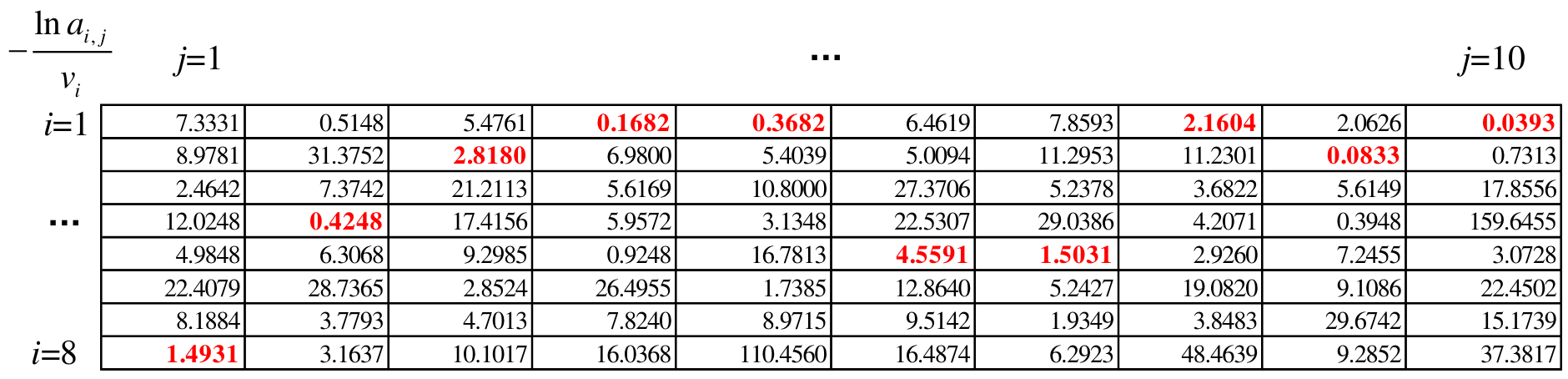}
	\caption{An example of computing $k$ independent Gumbel-Max variables $s_1,
    \ldots, s_k$ of a vector $\vec{v} =(0.3, 0.1, 0.05, 0.05, 0.2, 0.07, 0.1,
    0.03)$, where $k=10$.
		The Gumbel-Max variable $s_j$ equals the index of the smallest element (i.e.,
    the red and bold one) in the $j$-th column of matrix $\left[-\frac{\ln
        a_{i,j}}{v_i}\right]_{1\le i\le 8, 1\le j\le 10}$.}
	\label{fig:exmple}
\end{figure*}

% To formally define our problem,
% We formulate our problem.
% We first introduce some notations.
We first introduce some notations and then formulate our problem.
For a nonnegative vector $\vec{v}=(v_1,\ldots,v_n)$ and each element $v_i\geq
0$, let $\vec{v}^*=(v_1^*,\ldots,v_n^*)$ be the normalized vector of $\vec{v}$,
where
\[
  v_i^*\triangleq\frac{v_i}{\sum_{j=1}^n v_j},\quad i=1,\ldots,n.
\]
% When $\sum_{j=1}^n v_j=1$, we call $\vec{v}$ a normalized vector, that is,
% $\vec{v}=\vec{v}^*$.
Let $N^+_{\vec{v}}\triangleq\{i\colon v_i > 0, i=1,\ldots,n\}$ be the set of
indices of positive elements in $\vec{v}$, and $n^+_{\vec{v}}\triangleq
|N^+_{\vec{v}}|$ be its cardinality.

For each $i=1,\ldots,n$, we independently draw $k$ random samples $a_{i,1},\ldots,
a_{i, k}$ from the uniform distribution $\text{UNI}(0,1)$.
Note that $a_{i,1}, \ldots, a_{i,k}$ are the same for different vectors.
Given a nonnegative vector $\vec{v}$, we aim to fast compute its Gumbel-Max sketch
$\vec{s}(\vec{v}) = (s_1(\vec{v}), \ldots, s_k(\vec{v}))$, where
\begin{align*}
  s_j(\vec{v})
  &\triangleq\arg\max_{i\in N^+_{\vec{v}}} \ln v_i - \ln (- \ln a_{i,j})\\
  &\triangleq \arg\min_{i\in N^+_{\vec{v}}} -\frac{\ln a_{i,j}}{v_i}.
\end{align*}

To compute the Gumbel-Max sketches of a large collection of vectors
(e.g., bag-of-words representations of documents),
the straightforward method (also used in NodeSketch~\cite{yang2019nodesketch}) first instantiates variables $a_{i,1}, \ldots, a_{i,k}$ from $\text{UNI}(0,1)$ for each index $i=1, \ldots, n$.
Then, for each nonnegative vector $\vec{v}$, 
it enumerates each $i\in N^+_{\vec{v}}$ and compute $-\frac{\ln a_{i,1}}{v_i}, \ldots, -\frac{\ln a_{i,k}}{v_i}$.
The above method requires memory space $O(nk)$ to store all $\left[a_{i,j}\right]_{1\le i\le n, 1\le j\le k}$, 
and time complexity $O(k n^+_{\vec{v}})$ to obtain the Gumbel-Max sketch $\vec{s}(\vec{v})$ of each vector $\vec{v}$.
We note that $k$ is usually set to be hundreds or even thousands, 
therefore, the straightforward method costs a very large amount of memory space and time when the vector of interest has a large dimension, e.g., $n=10^9$.
To reduce the memory cost,
one can easily use hash techniques or random number generators with specific seeds (e.g., consistent random number generation methods in~\cite{ShrivastavaUAI2014,ShrivastavaICML2014,ShrivastavaICML2017}) to generate each of $a_{i,1}, \ldots, a_{i,k}$ on the fly,
which do not require to calculate and store variables $\left[a_{i,j}\right]_{1\le i\le n, 1\le j\le k}$ in memory.

To address the computational challenge, in this paper, we propose a method that reduces the time complexity of computing sketch $\vec{s}(\vec{v})$ from $O(k  n^+_{\vec{v}})$ to $O(k \ln k + n^+_{\vec{v}})$.
%A list of frequently used notations throughout the paper are given in Table~\ref{tab:notations}.
In the follows, when no confusion raises, we simply write $s_j(\vec{v})$ and $n^+_{\vec{v}}$ as $s_j$ and $n^+$ respectively.

%%% Local Variables:
%%% mode: latex
%%% TeX-master: "FastGM"
%%% End:

\section{Our Method}\label{sec:method}
In this section, we first introduce the basic idea behind our method FastGM.
Specially, we find that, for each element $v_i > 0$ in vector $\vec{v}$, the variables $-\frac{\ln a_{i, 1}} {v_i}, \ldots, -\frac{\ln a_{i, k}} {v_i}$ can be computed in ascending order and this procedure can be viewed as a Balls-and-Bins model.
Then, we derive our model BBM-Mix to randomly put balls one by one into $k$ empty bins, where each ball is assigned with a random variable in ascending order.
Based on BBM-Mix, we next introduce our method FastGM to compute the Gumbel-Max sketch $(s_1(\vec{v}), \ldots, s_k(\vec{v}))$ of vector $\vec{v}$.
Specifically, we model this procedure as randomly throwing balls arrived at different rates into $k$ empty bins and each bin records the timestamp of the first arrived ball.
%, where each ball is assigned with a random variable in ascending order.
When no bins are empty, we early stop the procedure,
and then each $s_j(\vec{v})$, $j=1, \ldots, k$, records the random variable of the first ball thrown into bin $j$.
At last, we discuss the time and space complexity of our method. 
%and also introduce the extension of our method FastGM to handle stream vectors of which elements arrive in a stream fashion.

\subsection{Basic Idea}
In Figure~\ref{fig:exmple}, we provide an example of generating a Gumbel-Max sketch of a vector $\vec{v} =(0.3, 0.1, 0.05, 0.05, 0.2,$ $0.07, 0.1, 0.03)$ to illustrate our basic idea, where we have $n=8$ and $k=10$.
Note that we aim to fast compute each $s_j = \arg\min_{1\le i\le 8} -\frac{\ln a_{i,j}}{v_i}$, $1\le j\le 10$.
i.e., the index of the minimum element in each column $j$ of matrix $\left[-\frac{\ln a_{i,j}}{v_i}\right]_{1\le i\le 8, 1\le j\le 10}$.
We generate matrix $\left[-\frac{\ln a_{i,j}}{v_i}\right]_{1\le i\le 8, 1\le j\le 10}$ based on the traditional Gumbel-Max Trick and mark the minimum element (i.e., the red and bold one indicating the Gumbel-Max variable) in each column $j$.
We find that Gumbel-Max variables tend to equal index $i$ with large weight $v_i$.
For example, among the values of all Gumbel-Max variables $s_1, \ldots, s_{10}$, index $1$ with $v_1 = 0.3$ appears 4 times, while index $3$ with $v_3 = 0.05$ never occurs.
Furthermore, let $R=25$, and $R_i = \lceil R v^*_i \rceil$, where $\vec{v}^*=(v_1^*, \ldots, v_8^*)$ is the normalized vector of $\vec{v}$.
We have $R_1=8$, $R_2 = 3$, $R_3 = 2$, $R_4 = 2$, $R_5 = 5$, $R_6 = 2$, $R_7 = 3$, and $R_8 = 1$.
%For matrix $\left[-\frac{\ln a_{i,j}}{v_i}\right]_{1\le i\le 8, 1\le j\le 10}$,
We also find that each Gumbel-Max variable occurs as one of a row $i$'s Top-$R_i$ minimal elements.
For example, the four Gumbel-Max variables occurring in the 1-st row are all among the Top-$R_1$ (i.e., Top-$8$) minimal elements.
Based on the above insights, we derive our method FastGM, of which elements in each row can be generated in ascending order.
As a result, we can early stop the computation when all the Gumbel-Max variables are acquired.
%Later in Section~\ref{subsec: BBN-MIX}, we show that elements in each row $i$ can be generated in ascending order.
Take Figure~\ref{fig:exmple} as an example, compared with the straightforward method that requires to compute all $nk=80$ random variables,
we compute $s_1, \ldots, s_k$ by only obtaining Top-$R_i$ minimal elements of each row $i$,
which significantly reduces the computation to around $\sum_{i=1}^8 R_i = 26$.

\subsection{A Building Block of Our Method FastGM}\label{subsec: BBN-MIX}
In this section, we introduce our model BBM-Mix to generate $k$ random variables in ascending order for each positive element $v_i$ of vector $\vec{v}$.
We define variables
\begin{equation}\label{eq:bij}
b_{i, j} = -\frac{\ln a_{i,j}}{v_i}, \quad j=1, \ldots, k.
\end{equation}
We easily observe that $b_{i, 1}, \ldots, b_{i, k}$
are equivalent to $k$ independent random variables generated according to the exponential distribution $\text{EXP}(v_i)$.
It is also well known that the first arrival time of a Poisson process with rate $v_i$ is a random variable following the exponential distribution $\text{EXP}(v_i)$.
Therefore, each $b_{i,j}$ can also be viewed as the first arrival time of a Poisson process $P_{ij}$ with rate $v_i$
and all Poisson processes $\{P_{i,j}: i=1,\ldots, n, j=1,\ldots,k\}$ are independent with each other.
Next, we show that $b_{i,j}$ can be generated via the following \emph{Balls-and-Bins Model} (in short, \emph{BBM}).

\subsubsection{Basic BBM}
In the basic BBM, balls arrive independently according to a Poisson process $\mathcal{P}_{i}$ with rate $kv_i$.
When a ball arrives at time $x$, we select a bin $j$ from $k$ bins at random, and then put the ball into bin $j$ as well as set $b_{i,j}= \min (b_{i,j}, x)$.
The register $b_{i,j}$ is used to record the timestamp of the first ball arriving in bin $j$.
When no urn is empty, we stop the process because all $b_{i, 1}, \ldots, b_{i, k}$ will not change anymore.
Then, we easily find that the arrival of balls at any bin $j$ is a Poisson process $P_{i,j}$ with rate $v_i$ because all balls are randomly split into $k$ bins and the Poisson process is splittable~\cite{MitzenmacherBook2005}.
The sequence of inter-arrival times of Poisson process $\mathcal{P}_{i}$ are independent and they are identically distributed exponential random variables with mean $\frac{1}{k v_i}$.
Therefore, the above BBM can be simulated as the following model.

\subsubsection{BBM-Hash}
We use a variable $x_i$ to record the time when the latest ball arrives.
Initialize $x_i=0$ and $b_{i,j} = +\infty$, $j=1,2,\ldots,k$ and then repeat the following Steps 1 to 3 until no bin is empty:
\begin{itemize}[leftmargin=1.1cm]
\item[\textbf{Step 1:}] Generate a random variable $u$ according to the uniform distribution $\text{UNI}(0,1)$;

\item[\textbf{Step 2:}] Compute $x_i = x_i - \frac{\ln u}{kv_i}$;

\item[\textbf{Step 3:}] Select a number $j$ from $\{1,\ldots, k\}$ at random, and then put a ball into bin $j$ as well as set $b_{i,j}= \min \{b_{i,j}, x_i\}$.
\end{itemize}

Clearly, it may require more than $k$ iterations to fill all these $k$ bins.
To compute the number of required iterations is exactly the coupon collector's problem (see~\cite{Cormen2001}, Chapter 5.4.2) and we find that $O(k \ln k)$ iterations are required in expectation.
To reduce the number of iterations, we propose another model BBM-Permutation.

\begin{algorithm}[t]
	\SetKwFunction{RandUNI}{RandUNI}
	\SetKwFunction{RandInt}{RandInt}
	\SetKwFunction{GetNextBalls}{\textbf{GetNextBalls}}
	\SetKwFunction{Return}{Return}
	\SetKwFunction{continue}{continue}
	\SetKwFunction{Swap}{Swap}
	\SetKwFunction{break}{break}
	\SetKwFunction{RandGamma}{RandGamma}
	\SetKwInOut{Input}{Input}
	\SetKwInOut{Output}{Output}
	\BlankLine
	\tcc{$v_i$ is the $i^{th}$ element of vector $\vec{v}$}
	\Input{$v_i$}
	\Output{$b_{i,1}, \ldots, b_{i,k}$}
	\BlankLine
	$x_i \gets 0$; $z_i \gets 0$; $m_i \gets k$\;
	$(\pi_{i, 1}, \ldots, \pi_{i, k})\gets (1, \ldots, k)$\;
	\While{$m_i> 0$}{
		$x_i, c \gets$ \GetNextBalls($i$)\;
		
		\If{$b_{i,c}$ is empty}{
			$b_{i,c}\gets \frac{x_i}{kv_i}$\;
		}
	}
	
	\caption{Pseudo code of BBM-Mix, where function $\text{GetNextBalls}(i)$ used in line 4 is defined in Algorithm~\ref{alg:getnextball}.\label{alg:bbmmix}}
\end{algorithm}

\begin{algorithm}[t]
	\SetKwFunction{RandUNI}{RandUNI}
	\SetKwFunction{RandInt}{RandInt}
	\SetKwFunction{GetNextBalls}{GetNextBalls}
	\SetKwFunction{Return}{Return}
	\SetKwFunction{continue}{continue}
	\SetKwFunction{Swap}{Swap}
	\SetKwFunction{break}{break}
	\SetKwFunction{RandGamma}{RandGamma}
	\SetKwInOut{Input}{Input}
	\SetKwInOut{Output}{Output}
	\BlankLine
	\tcc{$x_i$ is the timestamp of the last arrived ball and $c$ is the index of assigned bin. }
	\Input{$i$}
	\Output{$x_i$, $c$}
	\BlankLine

%	\tcc{The input $i$ of $\text{GetNextBalls}(i)$ indicates the $i$-th element of vector $\vec{v}$, and its outputs are $x_i$ and $c$, where $x_i$ is the timestamp of the last arrived ball when $v_i=\frac{1}{k}$, and $c$ is the assigned bin of these balls.}
%	\textbf{Function} $\textbf{GetNextBalls}(i$):\\
	\tcc{$z_i$ is the accumulated number of arrived balls and we use it to guarantee the consistency.}
	$seed\gets i||z_i$\;
%	\tcc{$\text{RandUNI}()$ returns a uniform random variable from the range $(0, 1)$.}
	$u \gets \RandUNI()$\;
	\If{$m_i > \phi_k$}{
		\tcc{$\text{RandInt}(k)$ returns a number from $\{1, 2 \ldots, k\}$ at random.}
		$j\gets \RandInt(k)$\;
		$x \gets -\ln u$\;
		$z \gets 1$\;
	}
	\Else{
		$z \gets \left\lfloor \frac{\ln u}{\ln (1-m_i/k)}\right\rfloor + 1$\;
		\tcc{$\text{RandGamma}(z, 1)$ returns a random variable that is gamma-distributed with shape $z$ and scale 1.}
		$x \gets \RandGamma (z, 1)$\;
		$j\gets \RandInt(m_i)$\;
	}
	\If{$j< m_i$}{
		$\Swap(\pi_{i, j}, \pi_{i, m_i})$\;
		$m_i\gets m_i - 1$\;
	}
	$x_i \gets x_i + x$\;
	$z_i \gets z_i + z$\;
	$c \gets \pi_{i, m_i}$\;
%	\Return $x_i$, $c$\;
	
	\caption{Pseudo code of GetNextBalls($i$).\label{alg:getnextball}}
\end{algorithm}

\subsubsection{BBM-Permutation}
For the above BBM-Hash, at Step 3, a nonempty bin $j$ may be selected and the value of $b_{i,j}$ will not change.
Therefore, BBM-Hash may need more than one iterations to encounter an empty bin especially when few bins are empty.
We use a variable $m_i$ to keep track of the number of empty bins and $k-m_i$ is the number of filled bins.
Let $z$ denote the number of iterations (balls) to encounter an empty bin.
We easily find that $z$ is a geometric random variable with success probability $\frac{m_i}{k}$ when $m_i<k$,
that is,
\[
P(z=l) = \left(1- \frac{m_i}{k}\right)^{l-1} \frac{m_i}{k}, \quad l=1, 2, \ldots.
\]
The results of these $z$ iterations is equivalent to the following procedure: Generate $z$ independent random variables $u_1, \ldots, u_z$ according to the uniform distribution $\text{UNI}(0, 1)$,
set $x_i = x_i - \sum_{l=1}^{z} \ln u_l$ at Step 2, and randomly select \emph{an empty bin} and set $b_{i,j} = \min \{b_{i,j}, x_i\}$ at Step 3.

Furthermore, because $-\ln u_1, \ldots, -\ln u_z$ are $z$ independent and identically distributed exponential random variables with mean 1,
the sum of these $z$ random variables, i.e., $- \sum_{l=1}^{z} \ln u_l$, is exactly a random variable that is distributed according to the Gamma distribution $\text{Gamma}(z, 1)$ with shape $z$ and scale 1.
Therefore, we can directly generate a random variable $x$ according to $\text{Gamma}(z, 1)$ and $x$ has the same probability distribution as $- \sum_{l=1}^{z} \ln u_l$.
This can significantly reduce the computational cost of generating the random variable $- \sum_{l=1}^{z} \ln u_l$ when $z$ is large.

Then, we derive the model BBM-Permutation, which outputs $b_{i,1}, \ldots, b_{i,k}$ with the same statistical distribution as BBM-Hash.
We use a vector $\vec{\pi}_i = (\pi_{i,1}, \ldots, \pi_{i,k})$ to record the index of empty and nonempty bins, which is initialized to $\vec{\pi}_i = (1, \ldots, k)$.
Specially, the first $m_i$ elements $\pi_{i,1}, \ldots, \pi_{i, m_i}$ are used to keep track of all remaining empty bins at the current time and the rest $k-m_i$ elements $\pi_{i,m_i +1}, \ldots, \pi_{i, k}$ are all current nonempty bins.
In addition, we initialize $m_i = k$ and $x_i=0$.
Then, we repeat the following Steps 1 to 5 until no bin is empty.
\begin{itemize}[leftmargin=1.1cm]
\item[\textbf{Step 1:}] Generate a random variable $u$ according to the uniform distribution $\text{UNI}(0,1)$;

\item[\textbf{Step 2:}] Set $z=1$ when $m_i=k$. Otherwise, we compute a variable $z = \left\lfloor \frac{\ln u}{\ln (1-m_i/k)}\right\rfloor + 1$,
which is a geometric random variable with success probability $\frac{m_i}{k}$.
Here we generate $z$ using the knowledge that $\left\lfloor \frac{\ln u}{\ln (1-p)}\right\rfloor + 1$ is a geometric random variable with success probability $p$~\cite{MitzenmacherBook2005};

\item[\textbf{Step 3:}] Generate a random variable $x$ according to Gamma distribution $\text{Gamma}(z, 1)$;

\item[\textbf{Step 4:}] Compute $x_i = x_i + \frac{x}{kv_i}$;

\item[\textbf{Step 5:}] Select a number $j$ from $\{1, \ldots, m_i\}$ at random, and then put $z$ balls into bin $\pi_{i, j}$ as well as set $m_i = m_i - 1$ and $b_{\pi_{i, j}}= x_i$.
In addition, we swap the values of $\pi_{i, j}$ and $\pi_{i, m_i}$.
\end{itemize}
Unlike BBM-Hash, more than one balls may occur at the same time
and all these balls are put into the same bin selected from the current empty bins at random.
BBM-Permutation requires exactly $k$ iterations to fill all bins.
In the end, $\vec{\pi}_i$ is exactly a random permutation of integers $1, \ldots, k$.
We design Step 5 inspired by the Fisher-Yates shuffle~\cite{Fisher1948},
which is a popular algorithm used for generating a random permutation of a finite sequence.
Compared with BBM-Hash using $64 k$ bits to store $b_{i,1}, \ldots, b_{i,k}$, BBM-Permutation also requires $k\log k$ bits to store $\pi_{i,1}, \ldots, \pi_{i,k}$.
Next, we elaborate our model BBM-Mix, which can be used to further accelerate the speed of BBM-Permutation.

\subsubsection{BBM-Mix}
Let $t_\text{H}$ denote the average computational cost required for one iteration of BBM-Hash.
For BBM-Hash, when there exist $m_i$ empty bins, we easily find that on average $\frac{k}{m_i}$ iterations are required to encounter the next empty bin.
Thus, the average computational cost of looking for and filling the next empty bin is $\frac{k t_\text{H}}{m_i}$ and the cost increases as the number of current empty bins $m_i$ decreases.
%When $m_i = 1$, one average $\frac{k}{m_i}$ iterations are required and the computational cost is $k t_\text{H}$.
We also let $t_\text{P}$ denote the average computational cost required for one iteration of BBM-Permutation.
For each iteration, BBM-Hash requires to generate two uniform random variables,
while BBM-Permutation requires to generate two uniform random variables and one Gamma random variable.
Therefore, BBM-Permutation requires more computations than BBM-Hash to complete an iteration,
i.e., $t_\text{P}$ is larger than $t_\text{H}$.
However, BBM-Permutation requires only one iteration to fill an empty bin selected at random.
%Its computational cost of looking for and filling the next empty bin is $t_\text{H}$.
As a result, we use a variable
\[
\phi_k = \frac{k t_\text{H}}{t_\text{P}},
\]
to determine whether BBM-Hash is faster than BBM-Permutation to fill an empty bin for a specific $m_i$.
Specifically, we use BBM-Hash when $m_i > \phi_k$ and otherwise BBM-Permutation. In our experiments, we find that $\frac{t_\text{H}}{t_\text{P}}\approx 10$, so we set $\phi_k = \frac{k}{10}$.

We can further accelerate the model by simplifying some operations.
Specially, we first reduce the division operations at BBM-Hash's Step 2 and BBM-Permutation's Step 4 by computing $x_i = x_i - \ln u$ and $x_i = x_i + x$
instead of $x_i = x_i - \frac{\ln u}{kv_i}$ and $x_i = x_i + \frac{x}{kv_i}$ respectively,
and at the end we enlarge $b_{i,1}, \ldots, b_{i, k}$ by $\frac{1}{kv_i}$ times.
Based on the above improvements, we derive our model BBM-Mix and the Pseudo code of BBM-Mix is given in Algorithm~\ref{alg:bbmmix}.
Initialize $\vec{\pi}_i = (\pi_{i,1}, \ldots, \pi_{i,k})= (1, \ldots, k)$, $m_i = k$, $x_i=0$, and $\phi_k = \frac{t_\text{H} k}{t_\text{P}}$. We run the following procedure until no bin is empty.
\begin{itemize}[leftmargin=1.1cm]
\item[\textbf{Step 1:}] Generate a random variable $u$ according to the uniform distribution $\text{UNI}(0,1)$.

\item[\textbf{Step 2:}] If $m_i > \phi_k$, go to Step 3.
Otherwise, go to Step 5;

\item[\textbf{Step 3:}] Compute $x_i = x_i - \ln u$;

\item[\textbf{Step 4:}] Select a number $j$ from set $\{1,\ldots, k\}$ at random.
If $j < m_i$ (i.e., now bin $\pi_{i, j}$ is empty), we put a ball into bin $\pi_{i, j}$, and set $b_{i, \pi_{i, j}}= x_i$.
In addition, we also swap the values of $\pi_{i, j}$ and $\pi_{i, m_i}$, and then set $m_i = m_i - 1$.
%Otherwise, we set $b_{i, \pi_{i, j}}= \min \{b_{i, \pi_{i, j}}, x_i\}$.
After this step, go to Step 1;

\item[\textbf{Step 5:}] Compute $z = \left\lfloor \frac{\ln u}{\ln (1-m_i/k)}\right\rfloor + 1$,
which is a geometric random variable with success probability $\frac{m_i}{k}$;

\item[\textbf{Step 6:}] Generate a random variable $x$ according to the Gamma distribution $\text{Gamma}(z, 1)$;

\item[\textbf{Step 7:}] Compute $x_i = x_i + x$;

\item[\textbf{Step 8:}] Select a number $j$ from $\{1, \ldots, m_i\}$ at random, and then put $z$ balls into bin $\pi_{i, j}$ as well as set $m_i = m_i - 1$ and $b_{\pi_{i, j}}= x_i$.
In addition, swap the values of $\pi_{i, j}$ and $\pi_{i, m_i}$.
When finishing this step, go to Step 1.
\end{itemize}
We easily find that BBM-Mix is faster than both BBM-Hash and BBM-Permutation while the generated variables $b_{i, 1}, \ldots, b_{i, k}$ also have the same statistical distribution as those of BBM-Hash and BBM-Permutation.
Last, we would like to point out that BBM-Mix has the same space complexity as BBM-Permutation.

\begin{algorithm}[t]
	\SetKwFunction{RandUNI}{RandUNI}
	\SetKwFunction{RandInt}{RandInt}
	\SetKwFunction{Return}{Return}
	\SetKwFunction{continue}{continue}
	\SetKwFunction{Swap}{Swap}
	\SetKwFunction{break}{break}
	\SetKwFunction{RandGamma}{RandGamma}
	\SetKwInOut{Input}{Input}
	\SetKwInOut{Output}{Output}
	\BlankLine
	
	\Input{$\vec{v} = (v_1, \ldots, v_n)$, $k$, $\phi_k$}
	\Output{$\vec{s} = (s_{1}, \ldots, s_{k})$}
	\BlankLine
	$R\gets 0$;    $k^*\gets k$;
	$(y_1, \ldots, y_k) \gets (-1, \ldots, -1)$\;
	\ForEach {$i\in N^+_{\vec{v}}$}{
		$x_i\gets 0$;  $z_i\gets 0$;      $m_i\gets k$\;
		$(\pi_{i,1}, \ldots, \pi_{i,k})\gets (1, \ldots, k)$\;
	}
	\tcc{The following part is LinearFill}
	\While {$k^*\neq 0$}{
		$R\gets R + \Delta$\;
		\ForEach {$i\in N^+_{\vec{v}}$}{
			$R_i\gets \lceil R v_i^*\rceil$\;
			\While{$m_i > 0$ and $z_i < R_i$}{
				$x_i, c \gets \textbf{GetNextBalls}(i)$\;
				$b_i\gets \frac{x_i}{k v_i}$\;
				\If{$y_c<0$}{
					$y_c\gets b_i$; $s_c\gets i$\; 
					$k^*\gets k^* - 1$\;
%					\continue\;
				}
				\ElseIf{$b_i< y_c$}{
					$y_c\gets b_i$; $s_c\gets i$\;
				}
			}
		}
	}
	\tcc{The following part is FastPrune}
	$j^* \gets \arg \max_{j=1,\ldots,k} y_j$\;
	$N\gets N^+_{\vec{v}}$\;
	\While {$N$ is not empty}{
		$R\gets R + \Delta$\;
		\ForEach {$i\in N$}{
			$R_i\gets \lceil R v_i^*\rceil$\;
			\While{$m_i > 0$ and $z_i < R_i$}{
				$x_i, c\gets \textbf{GetNextBalls}(i)$\;
				$b_i\gets \frac{x_i}{k v_i}$\;
				\If{$b_i > y_{j^*}$}{
					$N\gets N - \{i\}$\;
					\break\;
				}	
				\If{$b_i < y_c$}{
					$y_c\gets b_i$;              $s_c\gets i$\;
					\If {$c==j^*$}{
						$j^* \gets \arg \max_{j=1,\ldots,k} y_j$\;
					}
				}
			}
		}
	}
	\caption{Pseudo code of our FastGM, where function $\text{GetNextBalls}(i)$ used in lines 10 and 24 is defined in Algorithm~\ref{alg:getnextball}.\label{alg:FastGM-static}}
\end{algorithm}

\subsection{Our Method FastGM}
Based on the model BBM-Mix, generating a Gumbel-Max sketch of a vector can be equivalently viewed as
the procedure of throwing $kn^+_{\vec{v}}$ balls generated by $n^+_{\vec{v}}$ Poisson processes into $k$ bins, where each ball is assigned with a random variable in ascending order.
Before introducing our method FastGM in detail, we naturally have the following two fundamental questions for the design of FastGM:

\noindent\textbf{Question 1.} How to fast search balls with the smallest timestamps to fill bins from these $n^+_{\vec{v}}$ Poisson processes?

\noindent\textbf{Question 2.} How to early stop a Poisson process $\mathcal{P}_i$, $i\in N^+_{\vec{v}}$?

We first discuss Question 1.
We note that balls of different Poisson processes $\mathcal{P}_i$ arrive at different rates $kv_i$.
Recall the example in Figure \ref{fig:exmple}, the basic idea behind the following technique is that process $\mathcal{P}_i$ with high rate $kv_i$ is more likely to produce balls with the smallest timestamps (i.e. Gumbel-Max variables).
Specially, when $z$ balls have been generated, let $x_{i,z}$ denote the time of the latest ball occurred in our method BBM-Mix.
We easily find that $x_{i, z}$ can be represented as the sum of $z$ identically distributed exponential random variables with mean $\frac{1}{k v_i}$.
Therefore, the expectation and variance of variable $x_{i, z}$ are computed as
\begin{equation}\label{eq:xizexpvar}
\mathbb{E}(x_{i, z}) = \frac{z}{k v_i}, \quad \text{Var}(x_{i, z}) = \frac{z}{k^2 v_i^2}.
\end{equation}
We find that $\mathbb{E}(x_{i, z})$ is $l$ times smaller than $\mathbb{E}(x_{j, z})$ when $v_i$ is $l$ times larger than $v_j$.
To obtain the first $R$ balls of the joint of all Poisson processes $\mathcal{P}_i$, $i\in N^+_{\vec{v}}$, we let each process $\mathcal{P}_i$ generate
$
R_i = \lceil R v_i^* \rceil
$ balls.
Then, we have
$\sum_{i=1}^n R_i \approx R$.
For all $i\in N^+_{\vec{v}}$, their $x_{i, r_i}$ approximately have the same expectation.
\begin{equation}\label{eq:exiri}
\mathbb{E}(x_{i, R_i}\mid R) \approx \frac{R}{k \sum_{j=1}^n v_j}, \quad i\in N^+_{\vec{v}}.
\end{equation}
Therefore, the $R$ balls with the smallest timestamps are expected to be obtained.

Then, we discuss Question 2, which is inspired by the ascending-order random variables for each arrived ball.
We let each bin $j$ use two registers $y_j$ and $s_j$ to keep track of information of the ball with the smallest timestamp among its currently received balls,
where $y_j$ records the ball's timestamp and $s_j$ records the ball's origin, i.e., the ball is coming from Poisson process $\mathcal{P}_{s_j}$.
When all bins $1, \ldots, k$ are filled with at least one ball, we let $y^*$ keep track of the maximum value of $y_1, \ldots, y_k$, i.e.,
\[
y^* = \max_{j=1,\ldots,k} y_j.
\]
Then, we can stop Poisson process $\mathcal{P}_i$ when a ball getting from $\mathcal{P}_i$ has a timestamp the larger than $y^*$
because the timestamps of the subsequent balls from $\mathcal{P}_i$ are also larger than $y^*$, which will not change any $y_1, \ldots, y_k$ and $s_1, \ldots, s_k$.

Based on the above two discussions, we develop our final method FastGM to fast generate $k$ Gumbel-Max variables $\vec{s}_{\vec{v}}=(s_1, \ldots, s_k)$ of any nonnegative vector $\vec{v}$.
As shown in Algorithm~\ref{alg:FastGM-static}, FastGM consists of two modules: LinearFill and FastPrune.
LinearFill is designed to quickly search balls with smallest timestamps arriving from all processes $\mathcal{P}_1, \ldots, \mathcal{P}_n$ and fill all bins $1, \ldots, k$.
When no bin is empty, we start the FastPrune module to early stop each Poisson process $\mathcal{P}_i$, $i\in N^+_{\vec{v}}$.
We perform the procedure of FastPrune because
Poisson Process $\mathcal{P}_i$ may also get balls with timestamps
smaller than $y^*$ and the balls may change the values of $y_j$ and $s_j$ for some bins $j$ after the procedure of LinearFill.
Next, we introduce these two modules in detail.

\noindent\textbf{$\bullet$ LinearFill Module}: This module fast search balls with smallest timestamps, and consists of the following steps:
\begin{itemize}[leftmargin=1.1cm]
%\item[\textbf{Initiation Step:}]  Let the total number of balls $R=k$;

\item[\textbf{Step 1:}] Iterate on each $i\in N^+_{\vec{v}}$ and repeat function $\text{GetNextBall}(i)$ in Algorithm~\ref{alg:getnextball} until it has received not less than $\lceil R v_i^*\rceil$ balls since the beginning of the algorithm. Meanwhile, each bin $j$ uses registers $y_j$ and $s_j$ to keep track of information of its received ball having the smallest timestamp,
where $y_j$ records the ball's timestamp and $s_j$ records the index of the Poisson process where the ball comes from;

\item[\textbf{Step 2:}] If there exist any empty bins, we increase $R$ by $\Delta$ and then repeat Step 1.
Otherwise, we stop the LinearFill procedure.
\end{itemize}
For simplicity, we set the parameter $\Delta = k$.
In our experiments, we find that the value of $\Delta$ has a small effect on the performance of FastGM.

\noindent\textbf{$\bullet$  FastPrune Module}:
When all bins $1, \ldots, k$ have been filled by at least one ball.
We start the FastPrune module, which mainly consists of the following two steps:
\begin{itemize}[leftmargin=1.1cm]
\item[\textbf{Step 1.}] Compute $y^* = \max_{j=1,\ldots,k} y_j$.

\item[\textbf{Step 2.}] For each $i\in N^+_{\vec{v}}$, we repeat function $\text{GetNextBall}(i)$ in Algorithm~\ref{alg:getnextball} to generate balls.
When the ball has a timestamp larger than $y^*$, we terminate Poisson process $\mathcal{P}_i$.
Note that $y_1, \ldots, y_k$ and $s_1, \ldots, s_k$ are also updated by received balls at this step.
Therefore, $y^*$ may also decrease with the number of arriving balls,
which accelerates the speed of terminating all Poisson processes $\mathcal{P}_i$, $i\in N^+_{\vec{v}}$.
\end{itemize}

\subsection{Complexity}
%Next, we discuss the space and time complexity of our method FastGM.
%In addition, we also introduce our implementation to satisfy the consistency property~\cite{Manasse2010}.

\textbf{Space Complexity.} For a nonnegative vector $\vec{v}$ with $n^+_{\vec{v}}$ positive elements,
our method FastGM requires $k\log k$ bits to store $\vec{\pi}_i$ for each $i\in N^+_{\vec{v}}$,
and in summary, $n^+_{\vec{v}} k\log k$ bits are desired.
In addition, $64k$ bits are desired for storing $y_1, \ldots, y_k$ (we use 64-bit floating-point registers to record $y_1, \ldots, y_k$), and $k\log n$ bits are required for storing $s_1, \ldots, s_k$, where $n$ is the size of the vector. 
However, the additional memory is released immediately after computing the sketch, and is far smaller than the memory for storing the generated sketches of massive vectors (e.g. documents).
Therefore, FastGM requires $n^+_{\vec{v}} k\log k  +64k + k \log n$ bits when generating $k$ Gumbel-Max variables $\vec{s}(\vec{v}) = (s_1, \ldots, s_k)$ of $\vec{v}$.
%However, when processing large datasets with a great number of vectors, we can re-use $\vec{\pi}_i$
%However, we note that we can release the used memory space except $s_1, \ldots, s_k$ after we obtain $\vec{s}(\vec{v})$.
%Therefore, our method can process vectors one by one when dealing with a large number of vectors.
%In addition, we can also easily parallelize our method FastGM on a cluster of servers/cores and each server/core handles a subset of vectors.

\textbf{Time Complexity.}
%We derive the time complexity of our method from the traditional Gumbel-Max Trick.
We easily find that a nonnegative vector and its normalized vector
have the same Gumbel-Max sketch.
For simplicity, therefore we analyze the time complexity of our method for only normalized vectors.
Let $\vec{v}^* = (v^*_1, \ldots, v^*_n)$ be a normalized and nonnegative vector.
Define
\[
\tilde{y}_j = \min_{i\in N^+_{\vec{v}^*}} - \frac{\ln a_{i,j}}{v^*_i}, \quad j=1, \ldots, k,
\]
\[
\tilde{y}^* = \max_{j=1,\ldots,k} \tilde{y}_j.
\]
At the end of our FastPrune procedure, each register $y_j$ used in the procedure equals $\tilde{y}_j$ and register $y^*$ equals $\tilde{y}^*$.
Because $\frac{\ln a_{i,j}}{v^*_i} \sim \text{EXP}(v^*_i)$, we easily find that each $y_j$ follows the exponential distribution $\text{EXP}(\sum_{i=1}^{n} v^*_i) = \text{EXP}(1)$.
From~\cite{expectation}, we then easily have
\[
	\mathbb{E}(\tilde y^*) = \sum_{m=1}^k \frac{1}{m} = \ln k + \gamma, \quad \text{Var}(\tilde y^*) = \sum_{m=1}^k \frac{1}{m^2} < \sum_{m=1}^\infty \frac{1}{m^2} = \frac{\pi^2}{6},
\]
where $\gamma \approx 0.577$ is the Euler-Mascheroni constant.
From Chebyshev's inequality, we have
\[
P\left(|\tilde y^* - \mathbb{E}(\tilde y^*)| \ge \alpha \text{Var}(\tilde y^*) \right) \le \frac{1}{\alpha^2}.
\]
Therefore, $\tilde y^*\le \mathbb{E}(\tilde y^*) + \alpha \text{Var}(\tilde y^*)$ happens with a large probability when $\alpha$ is large. In other words, the random variable $\tilde y^*$ can be upper bounded by $\mathbb{E}(\tilde y^*) + \alpha \text{Var}(\tilde y^*)$ with a large probability.
Next, we derive the expectation of $x_{i,R}$ after the first $R$ balls generated by our method FastGM have been put into $k$ bins.
%Recall that register $y_j$ keeps track of the smallest timestamp of balls thrown in bin $j$.
For each Poisson process $\mathcal{P}_i$, $i\in N^+_{\vec{v}}$,
from equations~(\ref{eq:xizexpvar}) and~(\ref{eq:exiri}), we find that
its last produced ball among these first $R$ balls has a timestamp $x_{i, R_i}$ with the expectation
$\mathbb{E}(x_{i, R_i}\mid R)  \approx \frac{R}{k}$.
When $R=k (\mathbb{E}(\tilde y^*) + \alpha \text{Var}(\tilde y^*)) < k(\ln k + \gamma + \frac{\alpha \pi^2}{6})$, the probability of $\mathbb{E}(x_{i, R_i}) > \tilde y^*$ is almost 1 for large $\alpha$, e.g., $\alpha > 10$.
%Note that we have $\tilde y^* \le  y^*$ at any time.
Therefore, we find that after putting the first $O(k\ln k)$ balls into the $k$ bins, each Poisson process $\mathcal{P}_i$ is expected to be early terminated and so we are likely to acquire all the Gumbel-Max variables.
We also note that each positive element has to be enumerated in the FastPrune model, therefore, the total time complexity of our method FastGM is $O(k \ln k+n^+_{\vec{v}})$.

%\textbf{Consistency.} Our random number generators (i.e. RandInt() and RandGamma() in Algorithm~\ref{alg:getnextball}) are desired to be consistent across different vectors.
%To be more specific, the same element across different vectors should share the same random number sequence.
%We can easily achieve the required consistency by pre-generating the sequence of random numbers
%and storing them in memory.
%However, this method costs a large amount of memory space for high dimensional vectors.
%In this paper, we use random number generators with specific seeds to generate these random numbers on the fly.
%Specially, for each Poisson process $\mathcal{P}_i$, we use the pair of $i$ and $z_i$ as its seed.
% and by default we generate its desired random numbers using this seed to keep consistency between different vectors.
\section{Evaluation} \label{sec:results}
We evaluate our method FastGM with the state-of-the-arts on two tasks: \textbf{task 1}) probability Jaccard similarity estimation, and \textbf{task 2})  network embedding.
All algorithms run on a computer with a Quad-Core Intel(R) Xeon(R) CPU E3-1226 v3 CPU 3.30GHz processor.
To demonstrate the reproducibility of the experimental results, we make our source code publicly available\footnote{https://github.com/qyy0180/FastGM}.

\subsection{Dataset}
We conduct experiments on both real-world and synthetic datasets.
For task 1, we run experiments on six real-world datasets: Real-sim \cite{Wei2017Consistent}, Rcv1 \cite{Lewis2004RCV1}, Webspam \cite{Wang2012Evolutionary}, Libimseti \cite{konect:2017:libimseti}, Last.fm \cite{konect:2017:lastfm_song}, and MovieLens \cite{konect:2017:movielens-10m_rating}.
In detail, Real-sim \cite{Wei2017Consistent}, Rcv1 \cite{Lewis2004RCV1}, and Webspam \cite{Wang2012Evolutionary} are datasets of web documents from different resources, where each vector represents a document and each entry in the vector refers to the TF-IDF score of a specific word for the document.
Libimseti \cite{konect:2017:libimseti} is a dataset of ratings between users on the Czech dating site, where each vector refers to a user and each entry records the user's rating to another one.
Last.fm \cite{konect:2017:lastfm_song} is a dataset of listening history, where each vector represents a song and each entry in the vector is the number of times the song has been listened to by a specific user.
MovieLens \cite{konect:2017:movielens-10m_rating} is a dataset of movie ratings, where each vector is a user and each entry in the vector is that user's rating to a specific movie.

For task 2, we perform experiments on four real-world graphs: YouTube \cite{Tang2009Scalable}, Email-EU \cite{leskovec2007graph}, Twitter \cite{choudhury10}, and WikiTalk \cite{sun_2016_49561}. In detail, YouTube \cite{Tang2009Scalable} is a dataset of friendships between YouTube users. Email-EU \cite{leskovec2007graph} is an email communication network, where nodes represent individual persons and edges are the communications between persons. Twitter \cite{choudhury10} is the following network collected from twitter.com and WikiTalk \cite{sun_2016_49561} is the communication network of the Chinese Wikipedia.
The statistics of all the above datasets are summarized in Table~\ref{tab:datasets}.

\begin{table}[t]		
	\centering
	\caption{Statistics of all used datasets.}\label{tab:datasets}
	\begin{subtable}[t]{\linewidth}
		\centering	
		\caption{Datasets used in task 1\label{tab:datasets1}}
		\begin{tabular}{|c|c|c|} \hline
			\textbf{Dataset} &\textbf{\#Vectors}&\textbf{\#Features}\\ \hline
			Real-sim \cite{Wei2017Consistent}&72,309&20,958\\ \hline
			Rcv1 \cite{Lewis2004RCV1}&20,242&47,236\\ \hline
			Webspam \cite{Wang2012Evolutionary}&350,000&16,609,143\\ \hline
			Libimseti \cite{konect:2017:libimseti}&220,970&220,970\\ \hline
			Last.fm \cite{konect:2017:lastfm_song}&992&1,085,612 \\ \hline
			MovieLens \cite{konect:2017:movielens-10m_rating}&69,878& 80,555 \\ \hline
		\end{tabular}
		
	\end{subtable}
	\begin{subtable}[t]{\linewidth}
		\centering
		\caption{Datasets used in task 2\label{tab:datasets2}}
		\begin{tabular}{|c|c|c|} \hline
			\textbf{Dataset} &\textbf{\#Nodes}&\textbf{\#Edges}\\ \hline
			%			Blog \cite{Lei2009Relational}&10,312&333,983\\ \hline
			%			PPI \cite{Grover2016node2vec}&3,890&76,584\\ \hline
			%			Wiki \cite{Grover2016node2vec}&4,777&184,812\\ \hline
			YouTube \cite{Tang2009Scalable}&1,138,499&2,990,443\\ \hline
			Email-EU \cite{leskovec2007graph}&265,214&420,045\\ \hline
			Twitter \cite{choudhury10}&465,017&834,797\\ \hline
			WikiTalk \cite{sun_2016_49561}&1,219,241&2,284,546\\ \hline
		\end{tabular}	
	\end{subtable}
\end{table}
\subsection{Baseline}
For task 1, we compare our method with $\mathcal{P}$-MinHash \cite{moulton2018maximally} on probability Jaccard similarity estimation to evaluate the performance of FastGM.
To highlight the efficiency of FastGM, we further compare FastGM with the state-of-the-art weighted Jaccard similarity estimation method, BagMinHash \cite{ertl2018bagminhash}.
Notice that BagMinHash estimates a different metric and thus we only show results on efficiency.
For task 2, we compare our method with the state-of-the-art embedding algorithm NodeSketch \cite{yang2019nodesketch}.
Specially, we use FastGM to replace the module of generating Gumbel-Max sketches (i.e., node embeddings) in NodeSketch.
In our experiments, we vary the size of node embedding $k$ and set decay weight $a = 0.005$ and order of proximity $r = 5$ as suggested in the original paper of NodeSketch.

\subsection{Metric}
For task 1, we use the sketching time and root mean square error (RMSE) to measure the efficiency and the effectiveness respectively.
For task 2, besides the efficiency comparison with the original NodeSketch, we also evaluate the effectiveness of our method on two popular applications of graph embedding: node classification and link prediction.
Similar to~\cite{yang2019nodesketch}, we use the Macro and Micro F1 scores to measure the performance of node classification,
and Precision$@K$ and Recall$@K$ to evaluate the performance of link prediction, where Precision$@K$ (resp. Recall$@K$) is the precision (resp. recall) on top $K$ high-similarity testing node pairs.
All experimental results are empirically computed from 100 independent runs by default.

\begin{figure}[t]	
	\centering
	\begin{subfigure}[t]{0.49\linewidth}
		\centering
		\includegraphics[width=\linewidth]{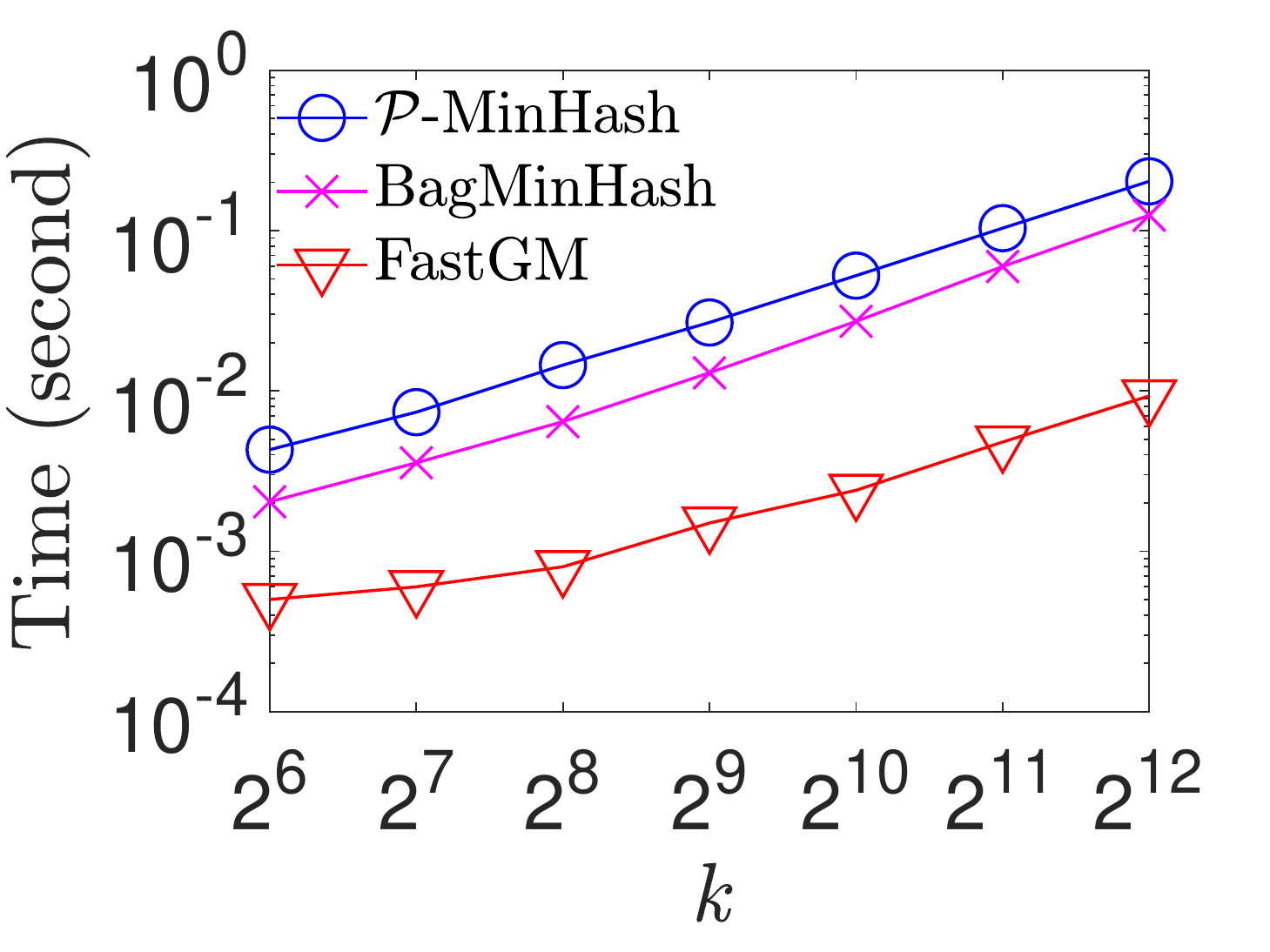}
		\caption{$n=10^3$}		
	\end{subfigure}
	\begin{subfigure}[t]{0.49\linewidth}
		\centering
		\includegraphics[width=\linewidth]{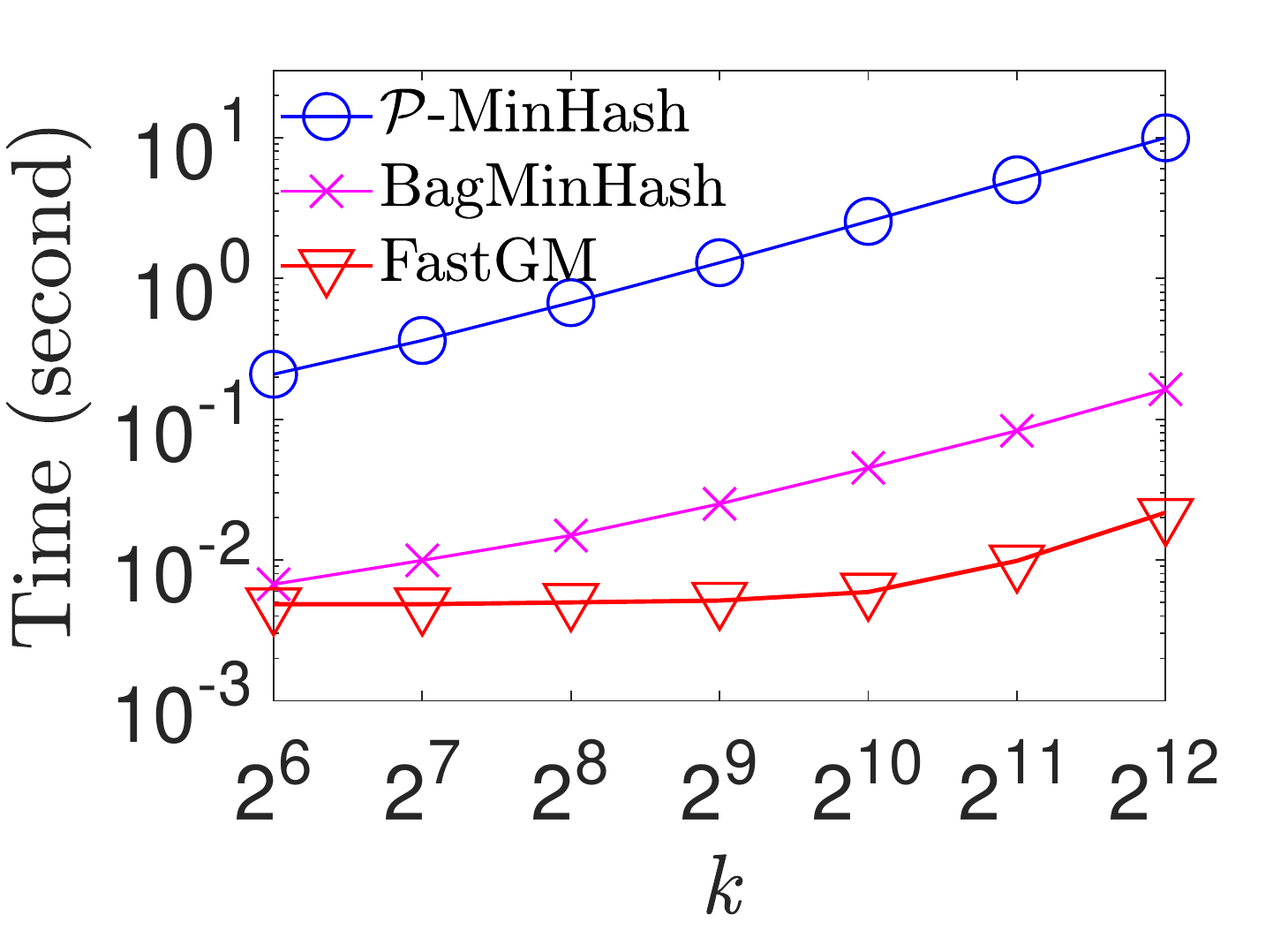}
		\caption{$n=10^4$}		
	\end{subfigure}
	\begin{subfigure}[t]{0.49\linewidth}
		\centering
		\includegraphics[width=\linewidth]{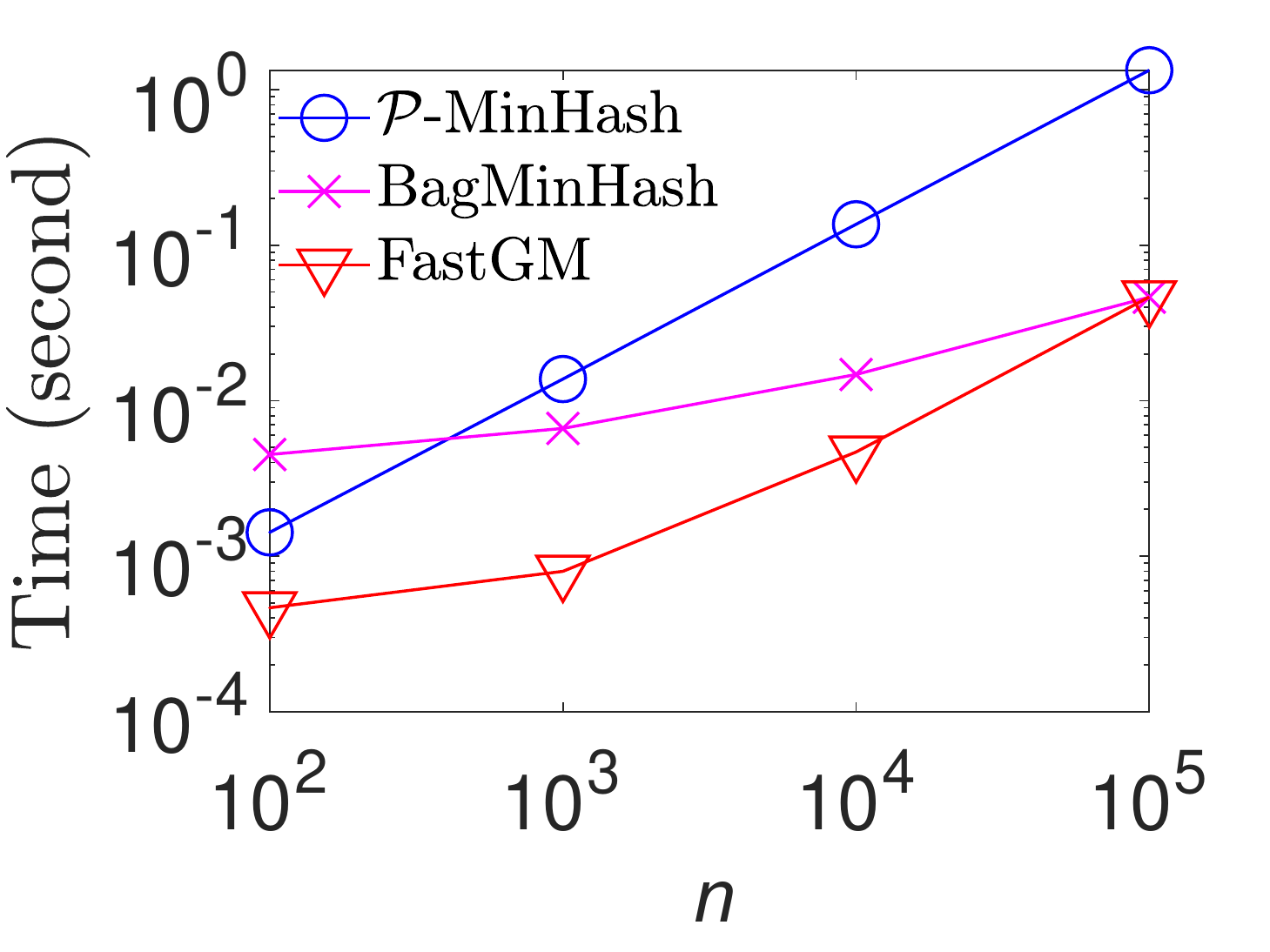}
		\caption{$k=2^8$}		
	\end{subfigure}
	\begin{subfigure}[t]{0.49\linewidth}
		\centering
		\includegraphics[width=\linewidth]{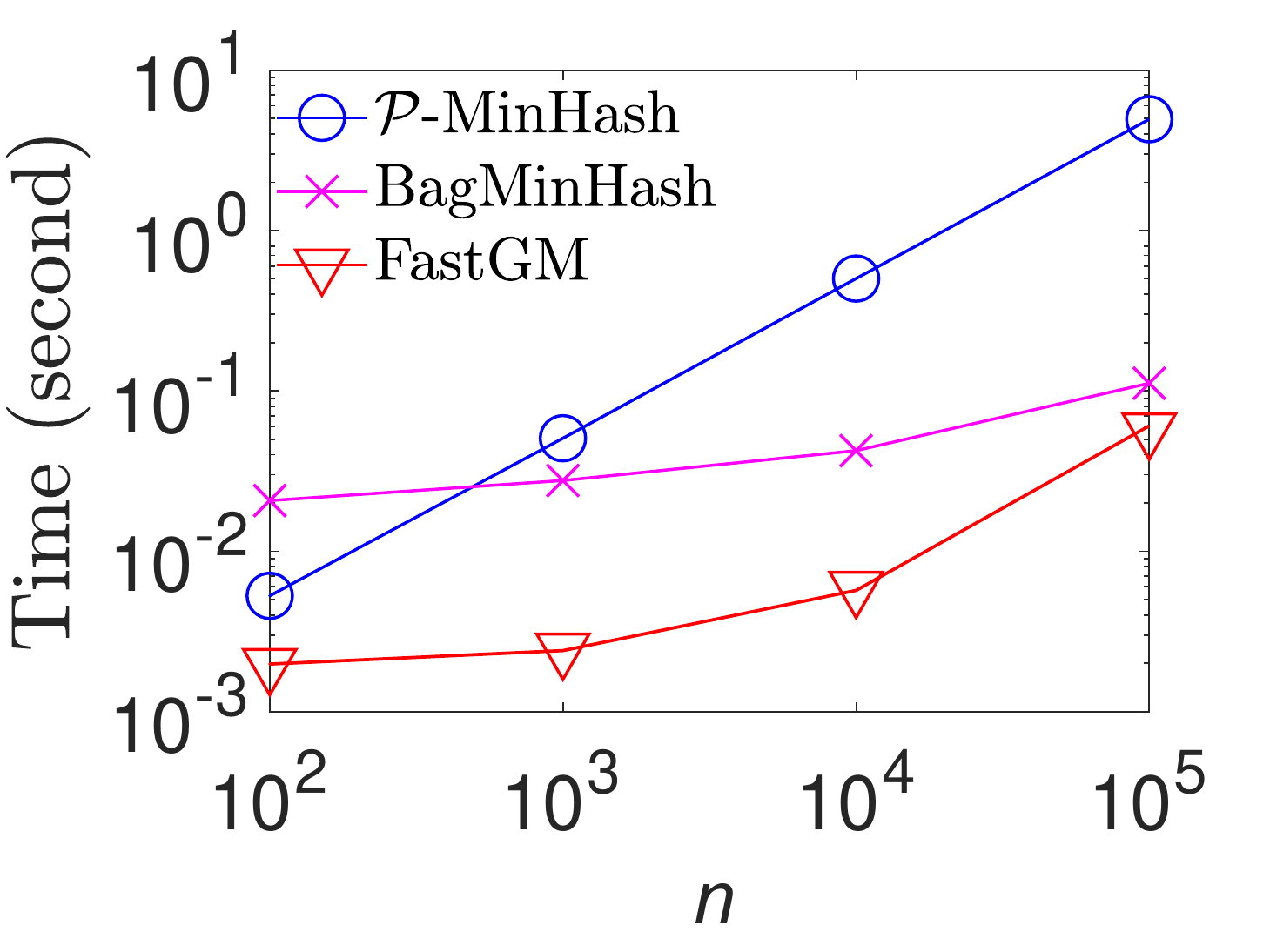}
		\caption{$k=2^{10}$}		
	\end{subfigure}
	\caption{(Task 1) Efficiency of FastGM compared with $\mathcal{P}$-MinHash and BagMinHash on synthetic vectors, where each element in the vector is randomly selected from UNI(0,1).}\label{fig:syndataset}
\end{figure}

\begin{figure}[t]
	\centering
	\begin{subfigure}[t]{0.49\linewidth}
		\centering
		\includegraphics[width=\linewidth]{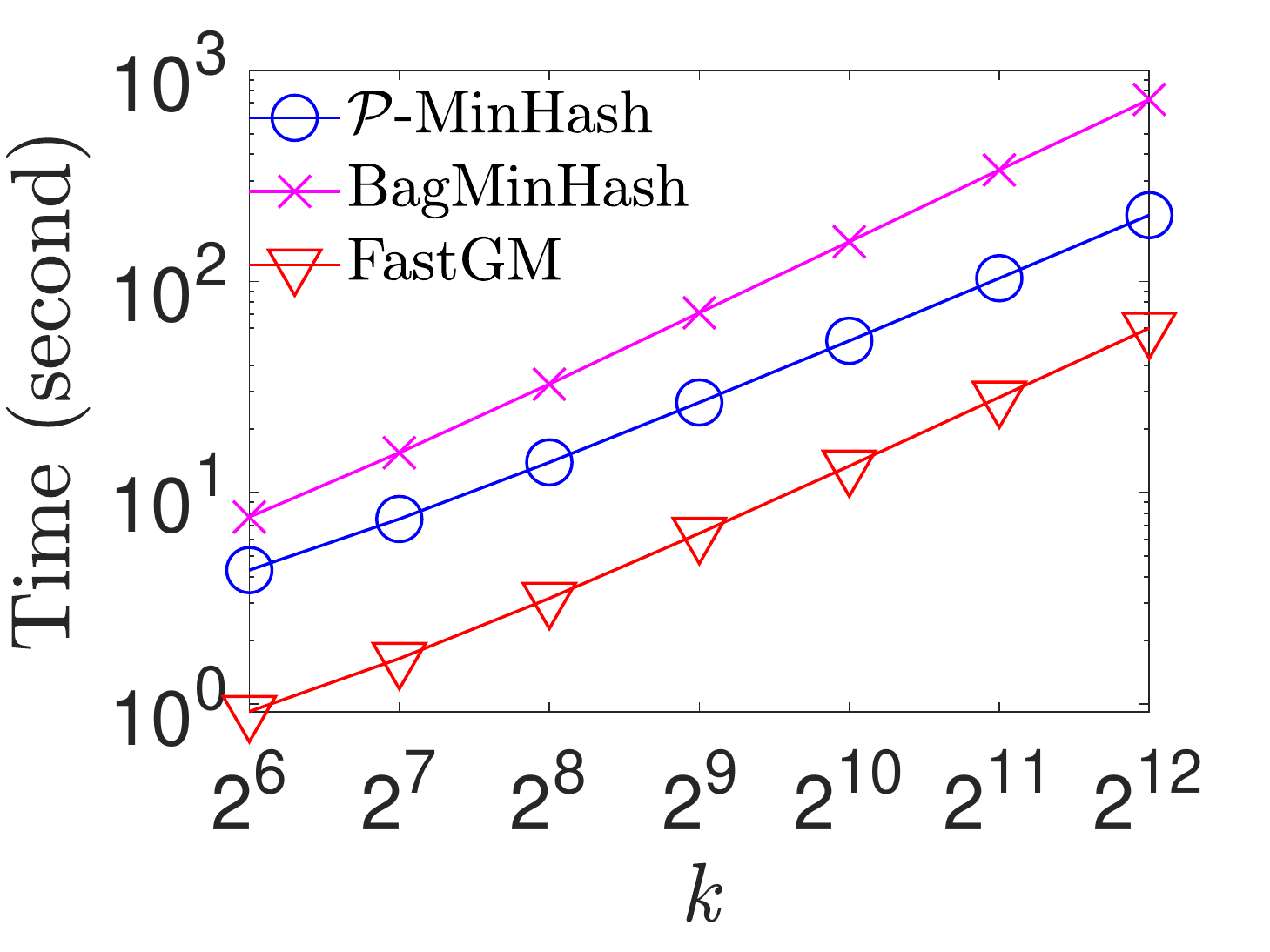}
		\caption{Real-sim}
	\end{subfigure}
	\begin{subfigure}[t]{0.49\linewidth}
		\centering
		\includegraphics[width=\linewidth]{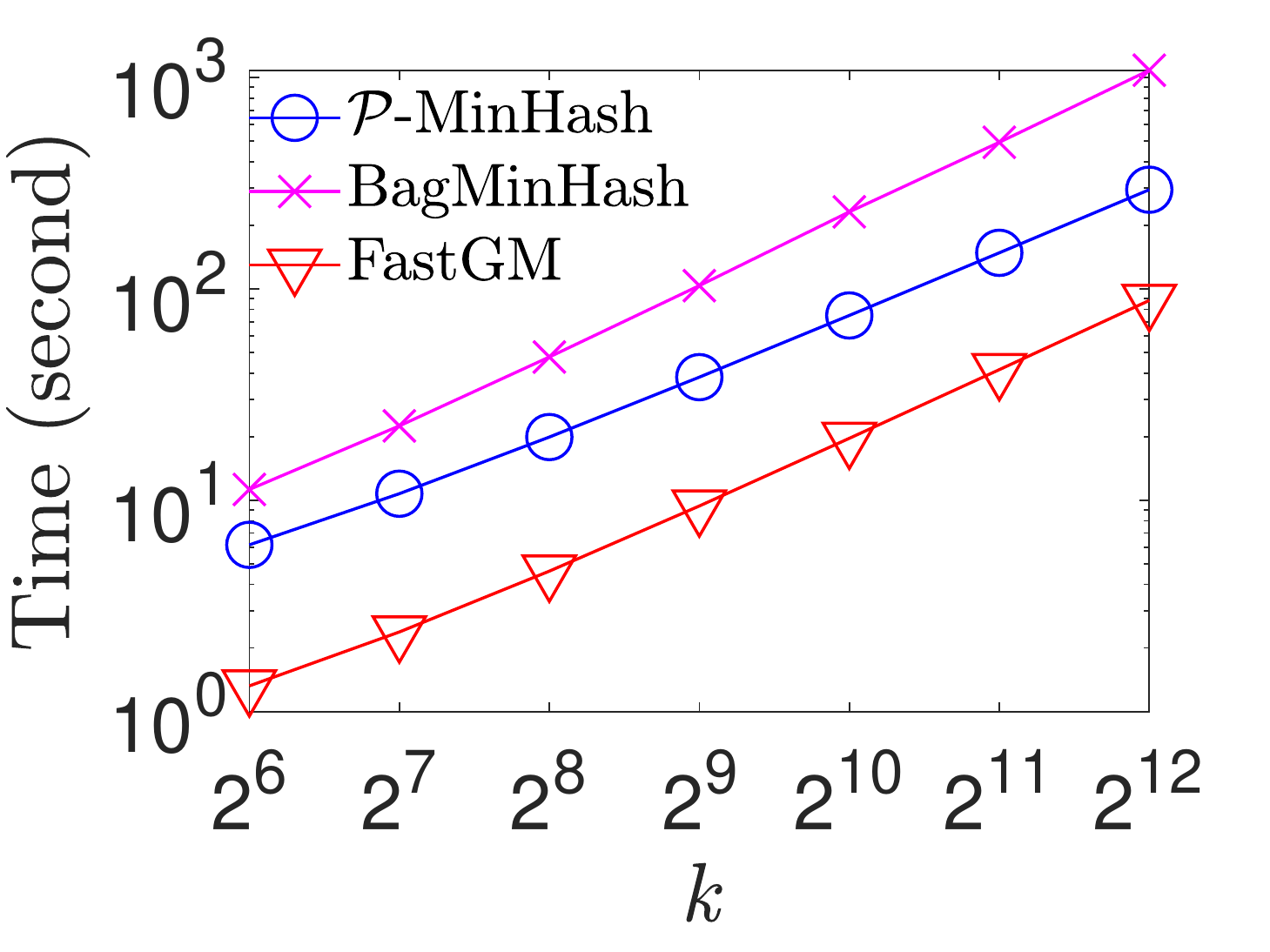}
		\caption{Rcv1}
	\end{subfigure}
	\begin{subfigure}[t]{0.49\linewidth}
		\centering
		\includegraphics[width=\linewidth]{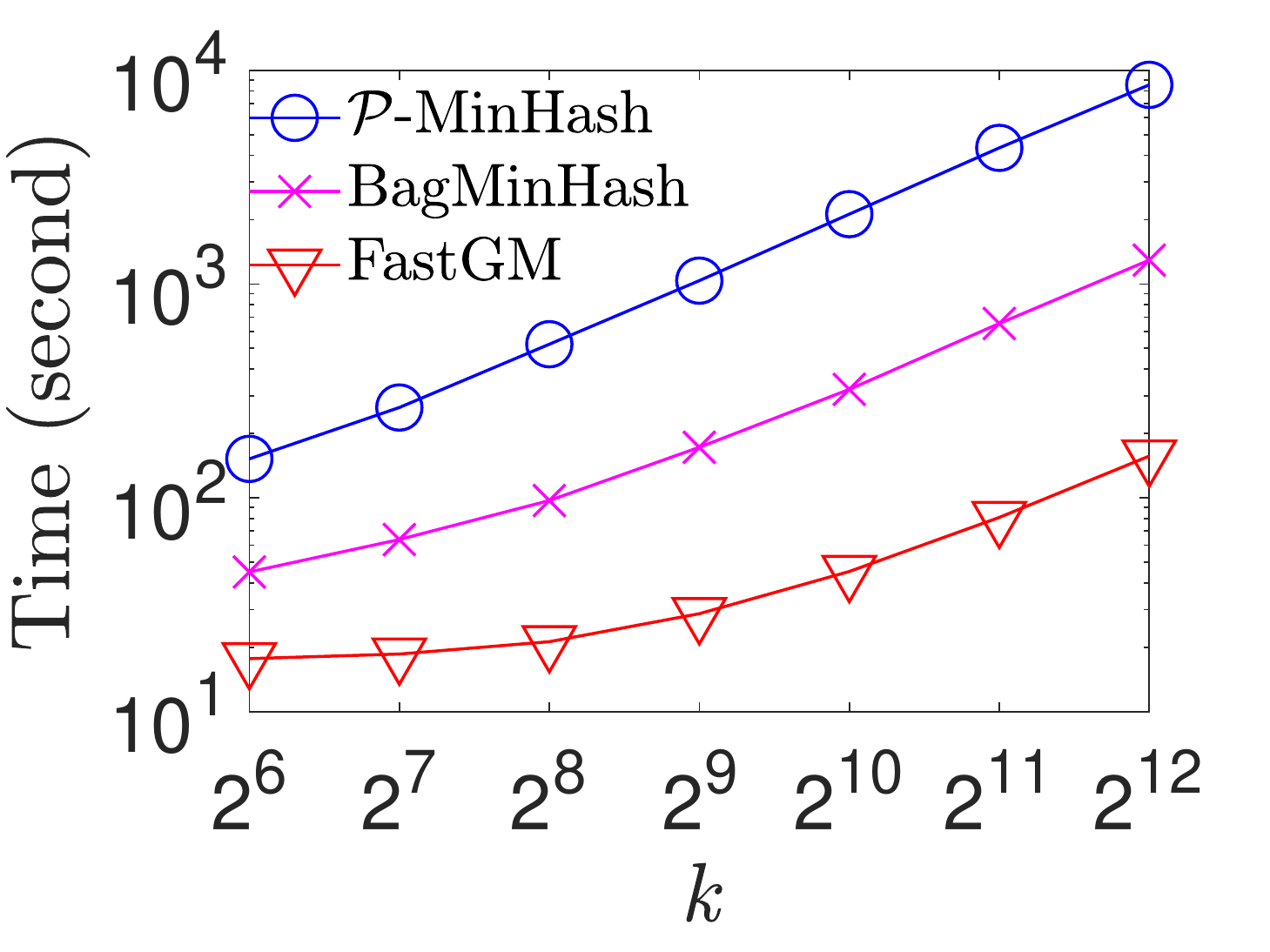}
		\caption{Webspam}
	\end{subfigure}	
	\begin{subfigure}[t]{0.49\linewidth}
		\centering
		\includegraphics[width=\linewidth]{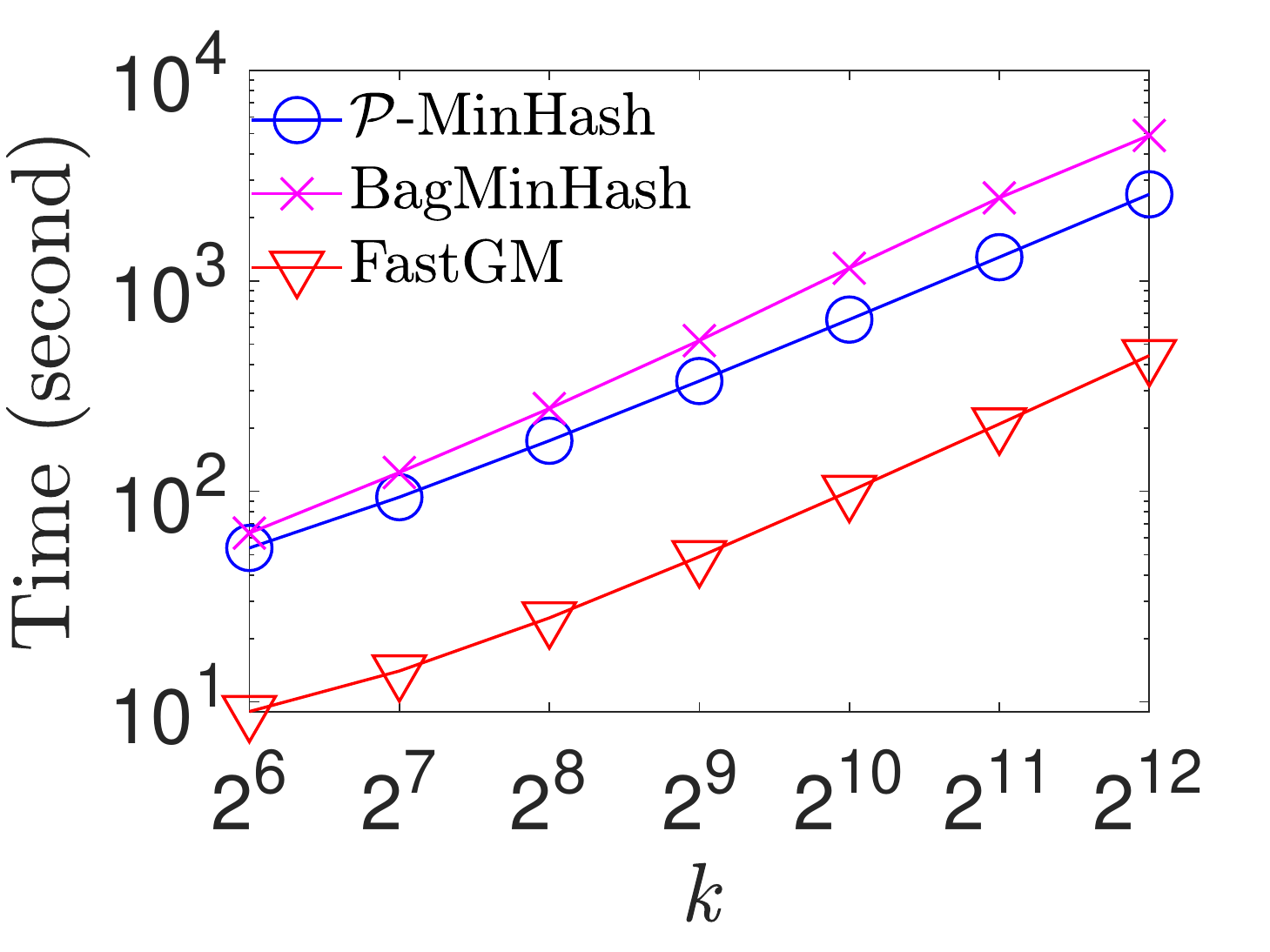}
		\caption{Libimseti}
	\end{subfigure}
	\begin{subfigure}[t]{0.49\linewidth}
		\centering
		\includegraphics[width=\linewidth]{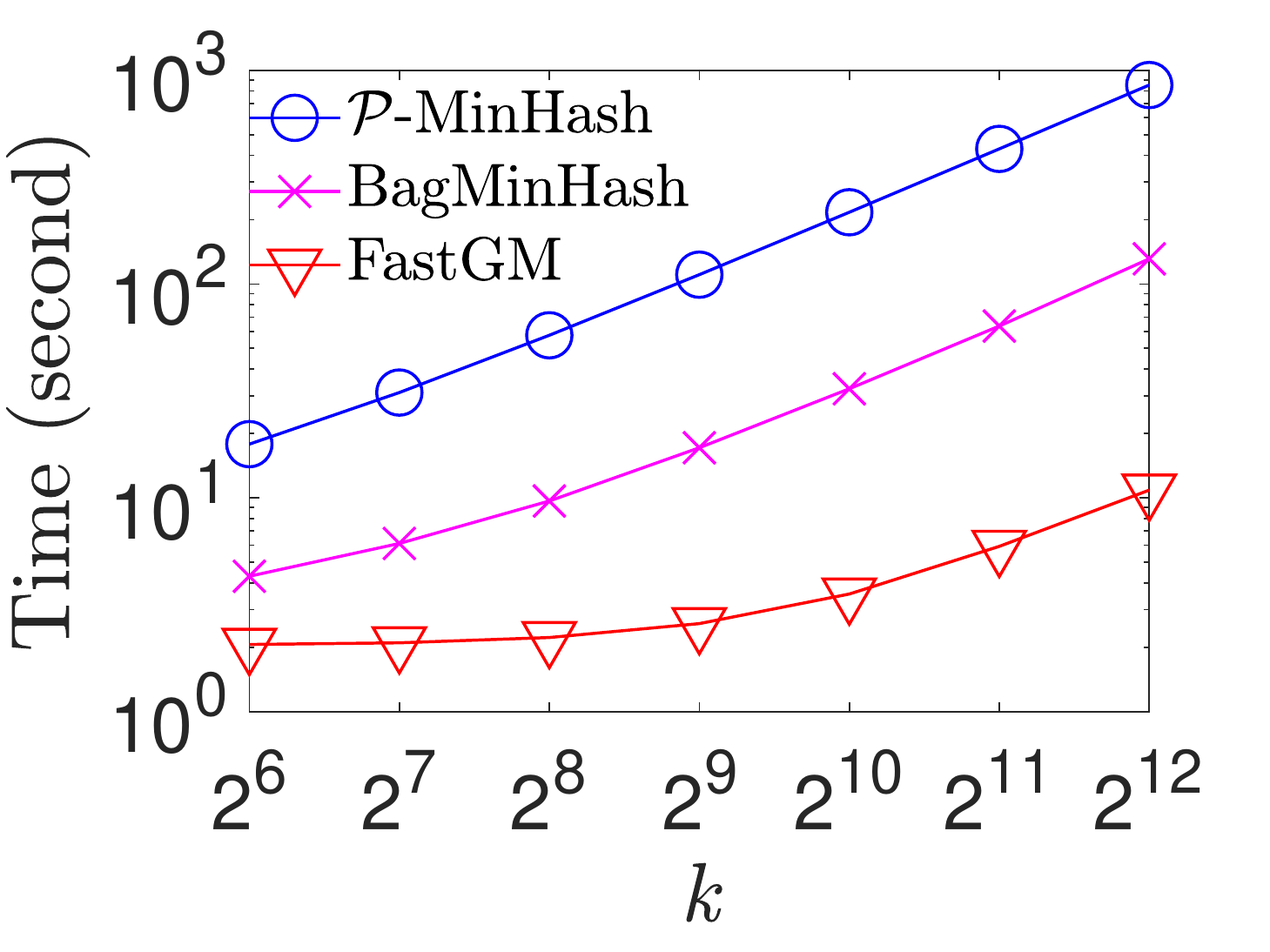}
		\caption{Last.fm}
	\end{subfigure}
	\begin{subfigure}[t]{0.49\linewidth}
		\centering
		\includegraphics[width=\linewidth]{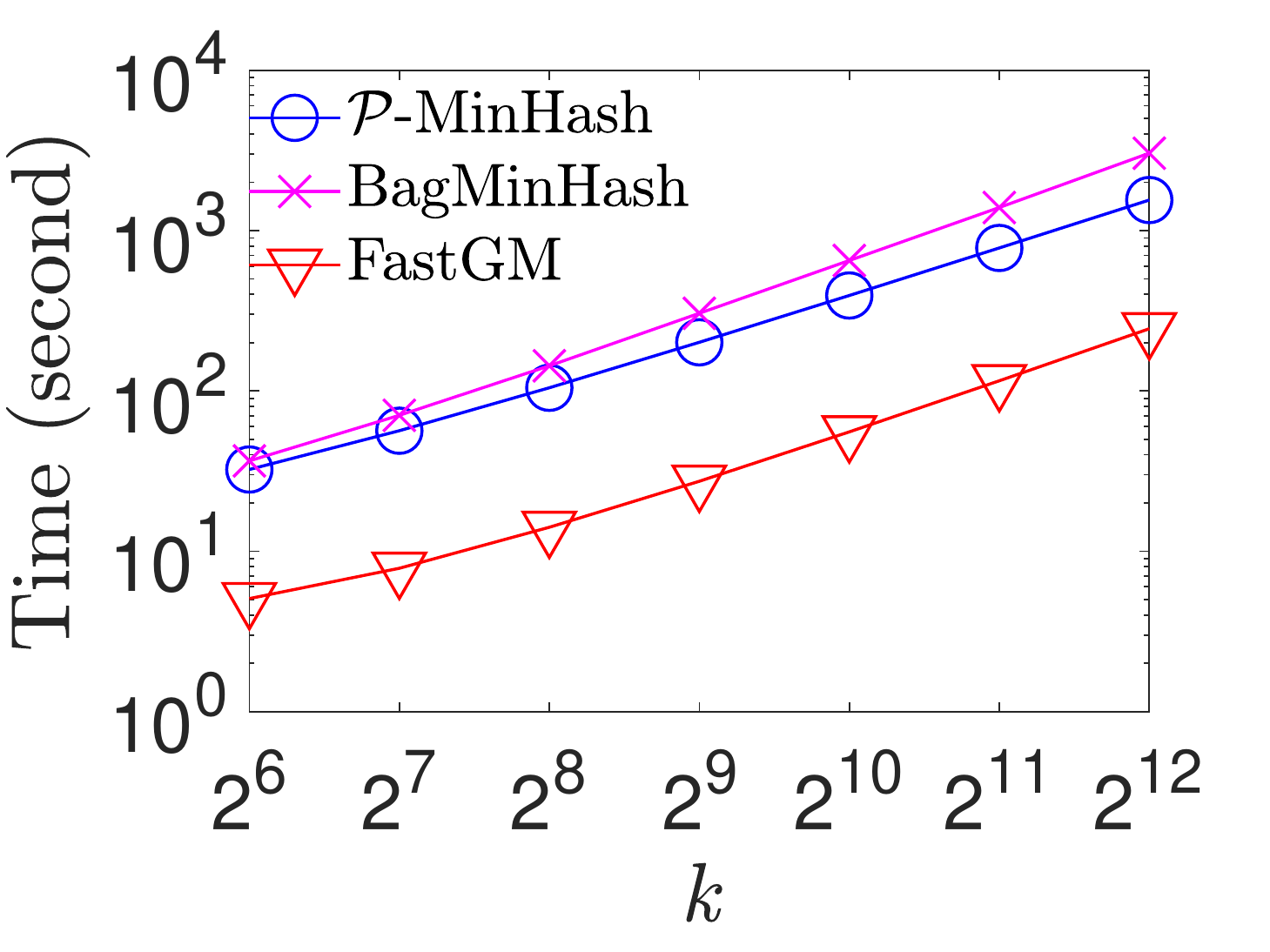}
		\caption{MovieLens}
	\end{subfigure}
	\caption{(Task 1) Efficiency of FastGM compared with $\mathcal{P}$-MinHash for different $k$.}\label{fig:time_realdata}
\end{figure}

\begin{figure}[t]
	\centering
	\begin{subfigure}[t]{0.49\linewidth}
		\centering
		\includegraphics[width=\linewidth]{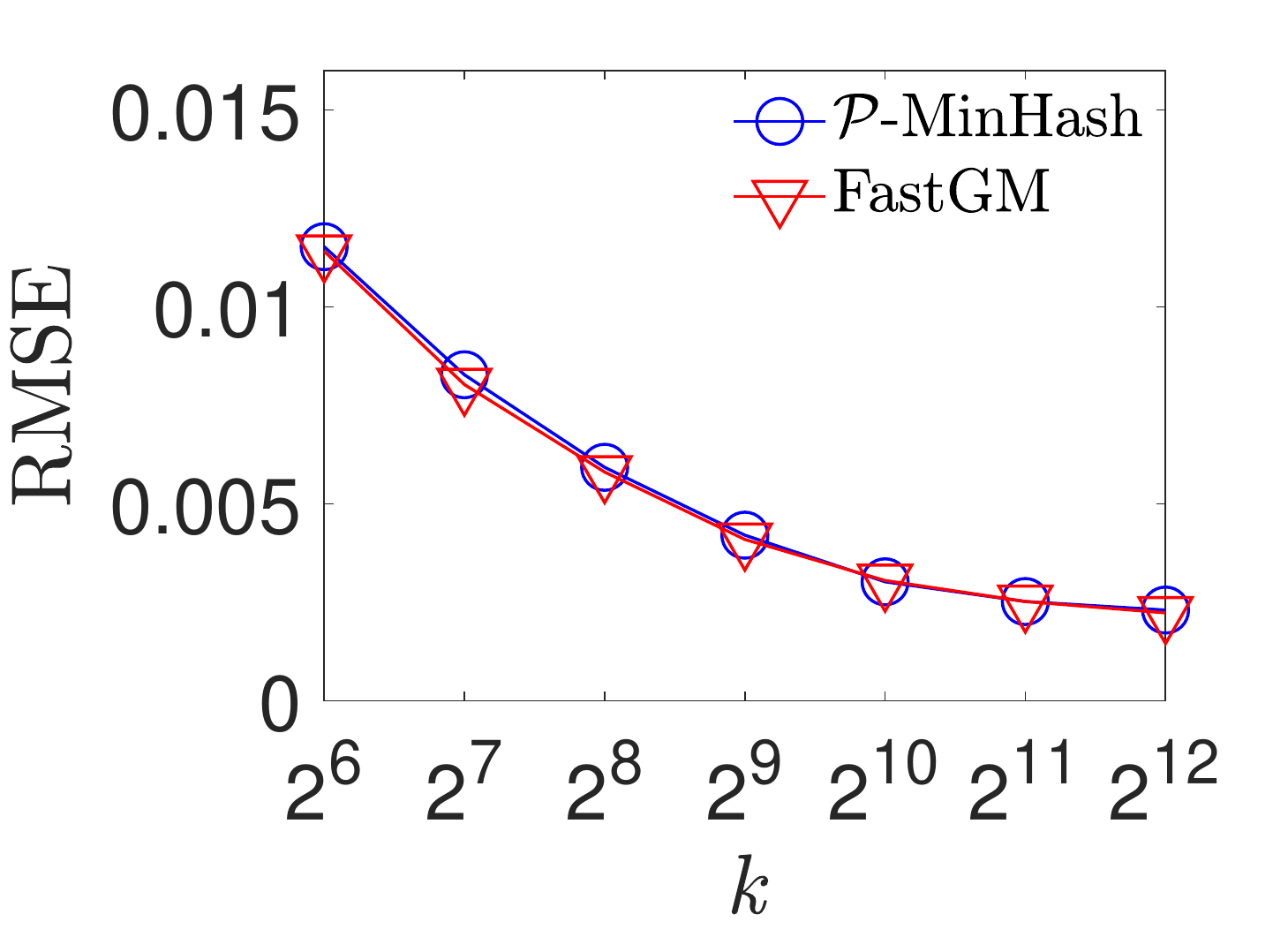}
		\caption{Real-sim}
	\end{subfigure}
	\begin{subfigure}[t]{0.49\linewidth}
		\centering
		\includegraphics[width=\linewidth]{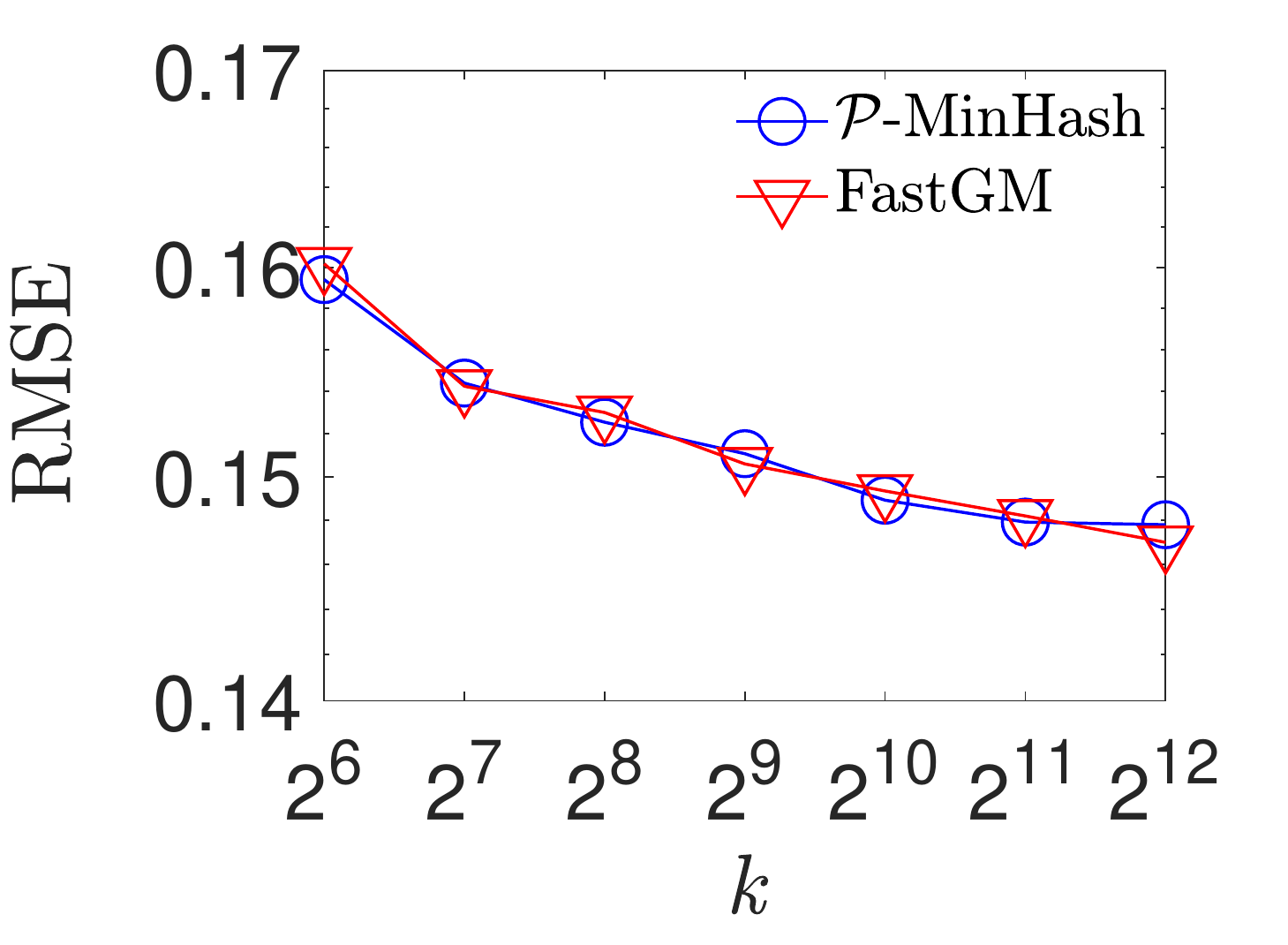}
		\caption{Webspam}
	\end{subfigure}
	\caption{(Task 1) Accuracy of FastGM compared with $\mathcal{P}$-MinHash for different $k$.}\label{fig:rmse}
\end{figure}

\begin{figure}[t]	
	\centering
	\begin{subfigure}[t]{0.49\linewidth}
		\centering
		\includegraphics[width=\linewidth]{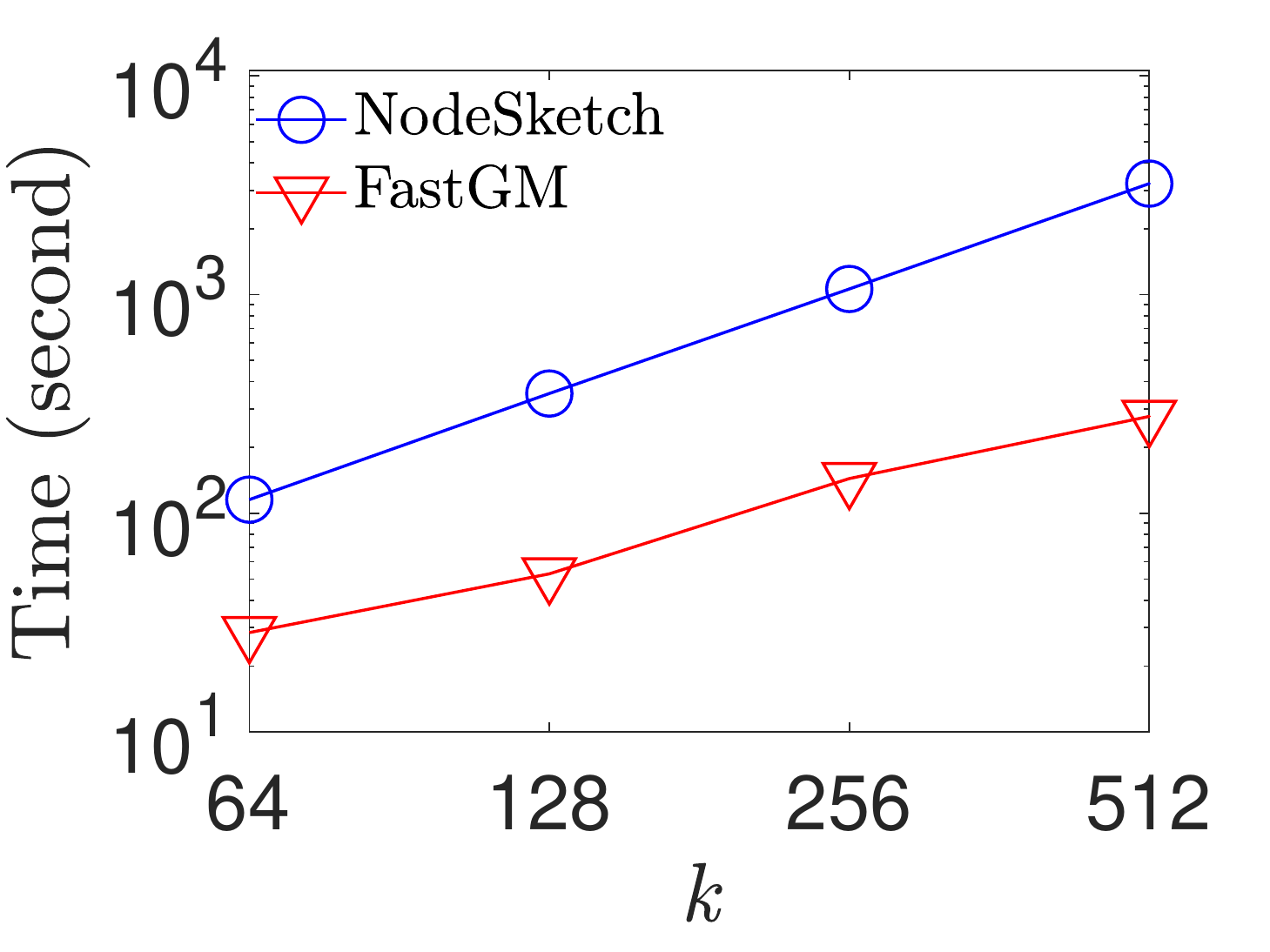}
		\caption{YouTube, sketching time}		
	\end{subfigure}
	\begin{subfigure}[t]{0.49\linewidth}
		\centering
		\includegraphics[width=\linewidth]{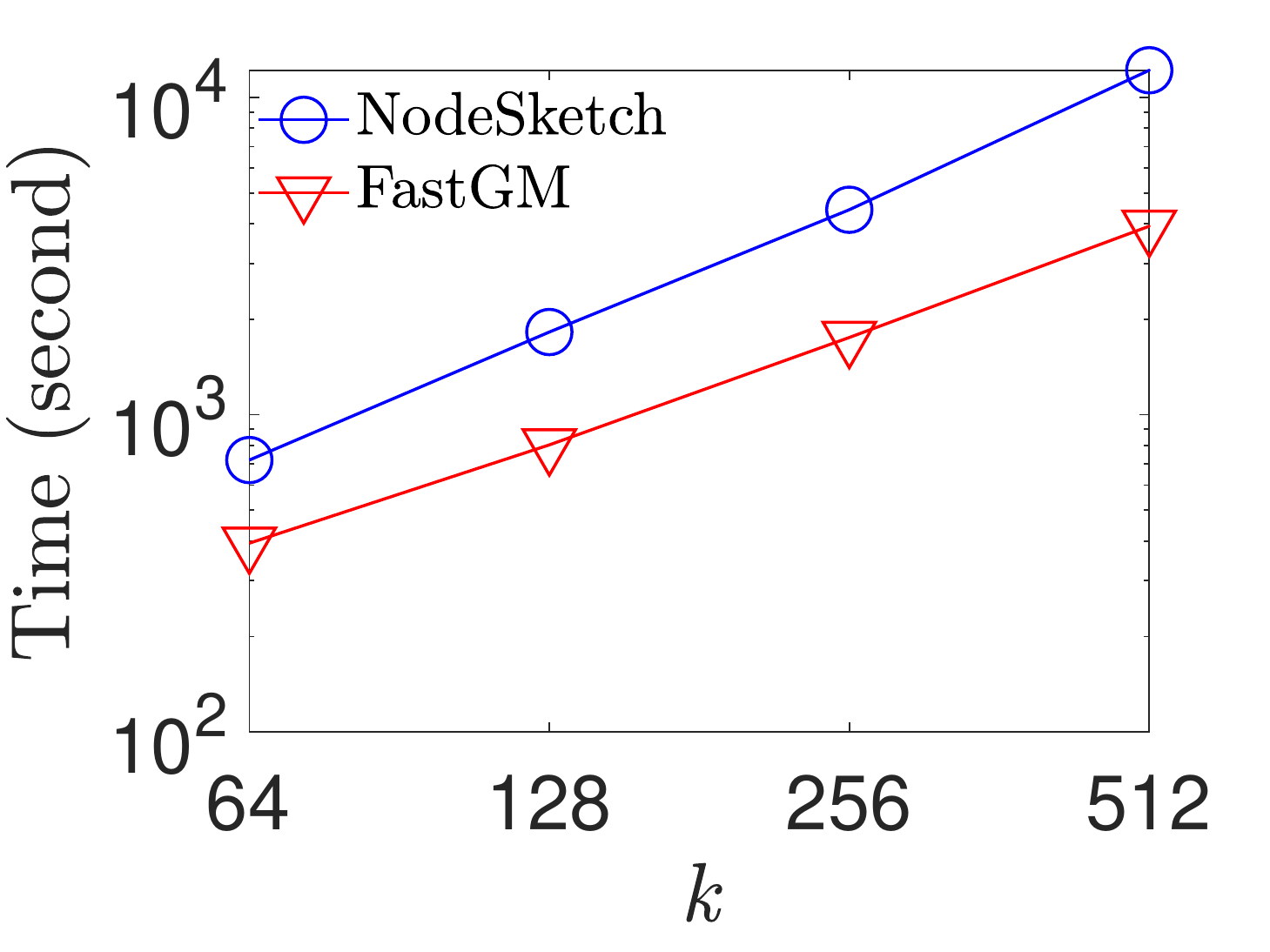}
		\caption{YouTube, total time}		
	\end{subfigure}
	%	\begin{subfigure}[t]{0.49\linewidth}
	%		\centering
	%		\includegraphics[width=\linewidth]{figs/IMDB_S}
	%		\caption{IMDB, embedding time}
	%	\end{subfigure}
	%	\begin{subfigure}[t]{0.49\linewidth}
	%	\centering
	%	\includegraphics[width=\linewidth]{figs/IMDB_T}
	%	\caption{IMDB, merging time}
	%	\end{subfigure}
	%	\begin{subfigure}[t]{0.49\linewidth}
	%		\centering
	%		\includegraphics[width=\linewidth]{figs/Douban_S}
	%		\caption{Douban, embedding time}		
	%	\end{subfigure}
	%	\begin{subfigure}[t]{0.49\linewidth}
	%	\centering
	%	\includegraphics[width=\linewidth]{figs/Douban_T}
	%	\caption{Douban, merging time}		
	%	\end{subfigure}
	\begin{subfigure}[t]{0.49\linewidth}
		\centering
		\includegraphics[width=\linewidth]{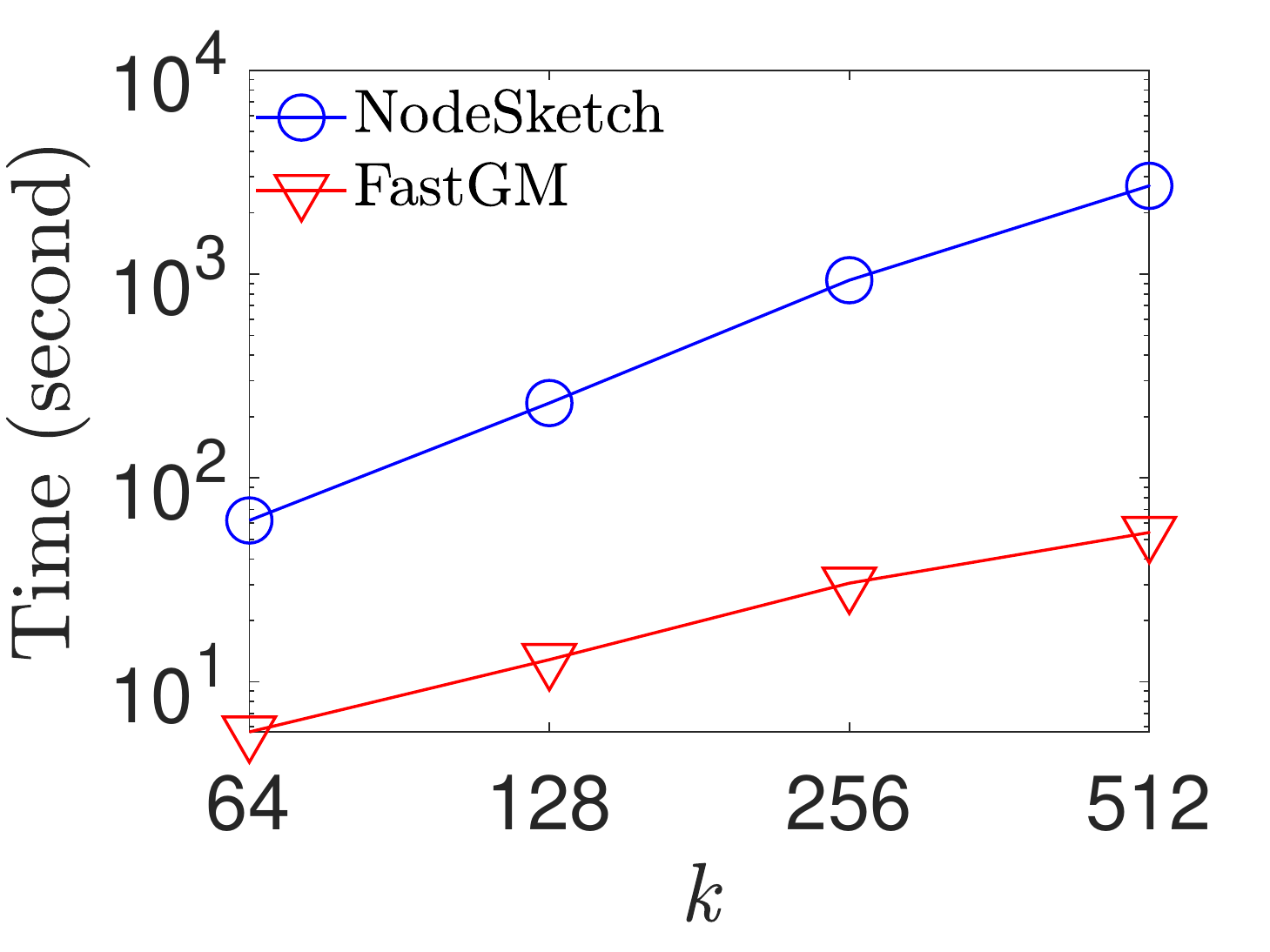}
		\caption{Email-EU, sketching time}		
	\end{subfigure}
	\begin{subfigure}[t]{0.49\linewidth}
		\centering
		\includegraphics[width=\linewidth]{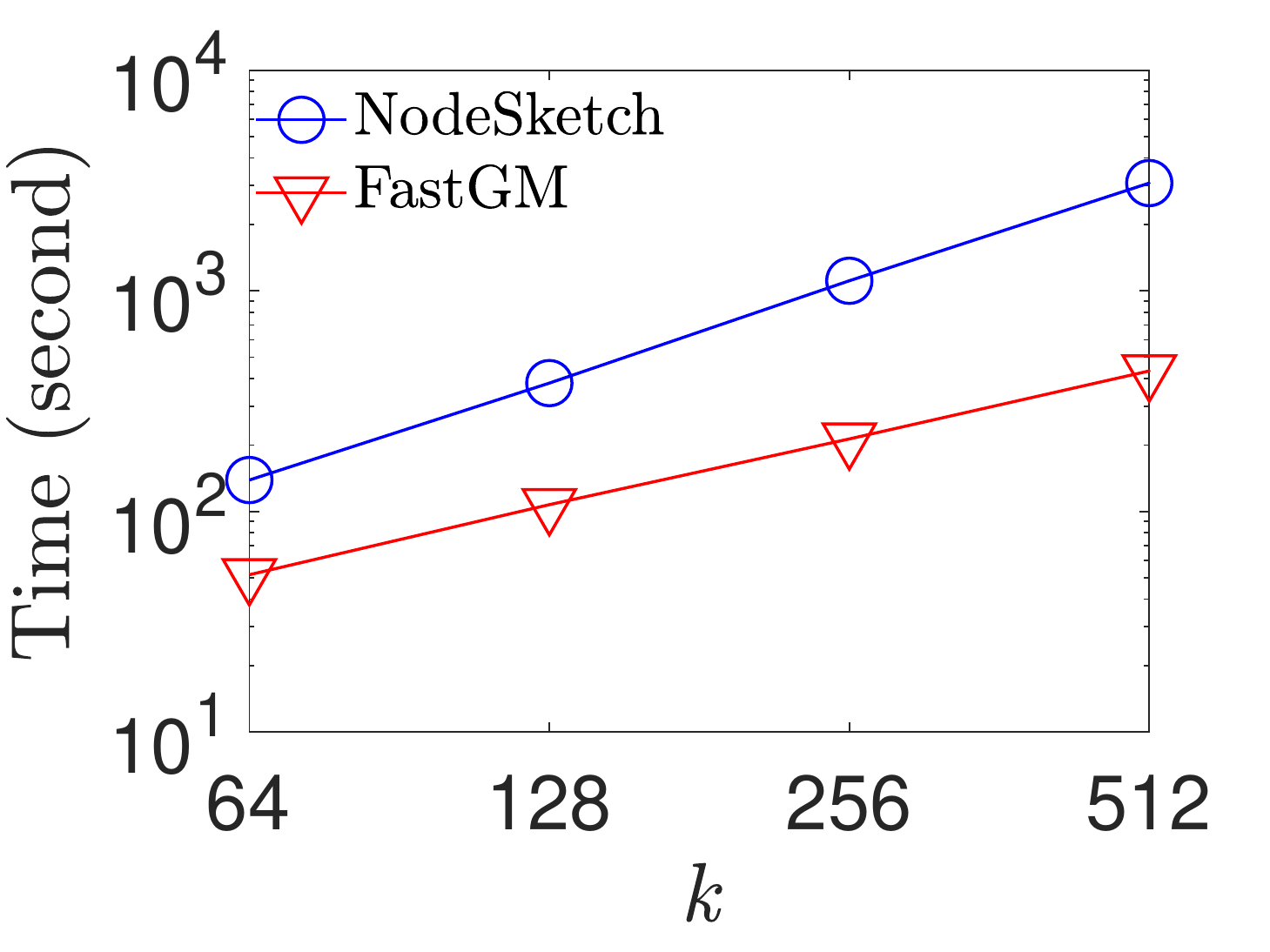}
		\caption{Email-EU, total time}		
	\end{subfigure}
	\begin{subfigure}[t]{0.49\linewidth}
		\centering
		\includegraphics[width=\linewidth]{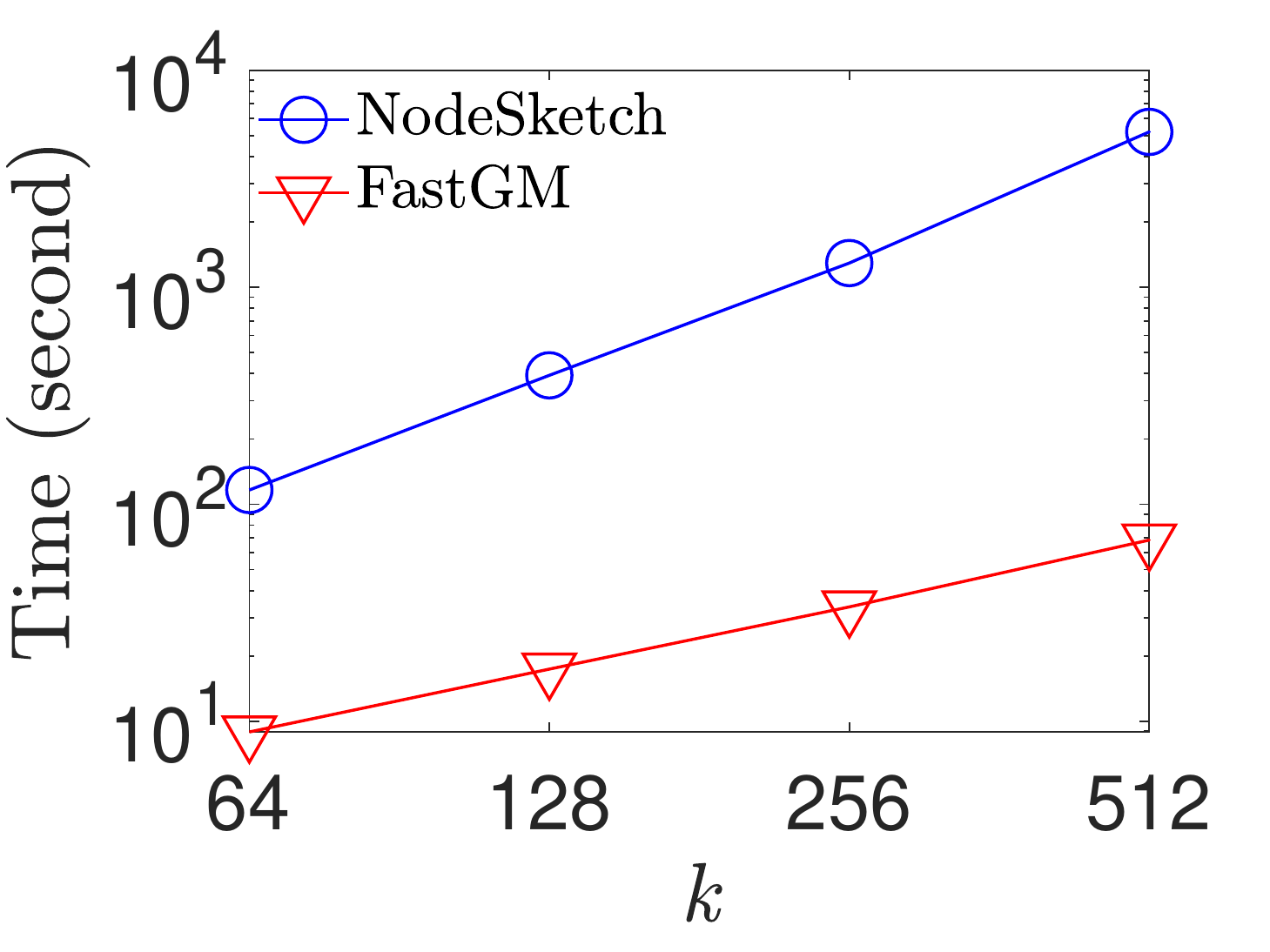}
		\caption{Twitter, sketching time}		
	\end{subfigure}
	\begin{subfigure}[t]{0.49\linewidth}
		\centering
		\includegraphics[width=\linewidth]{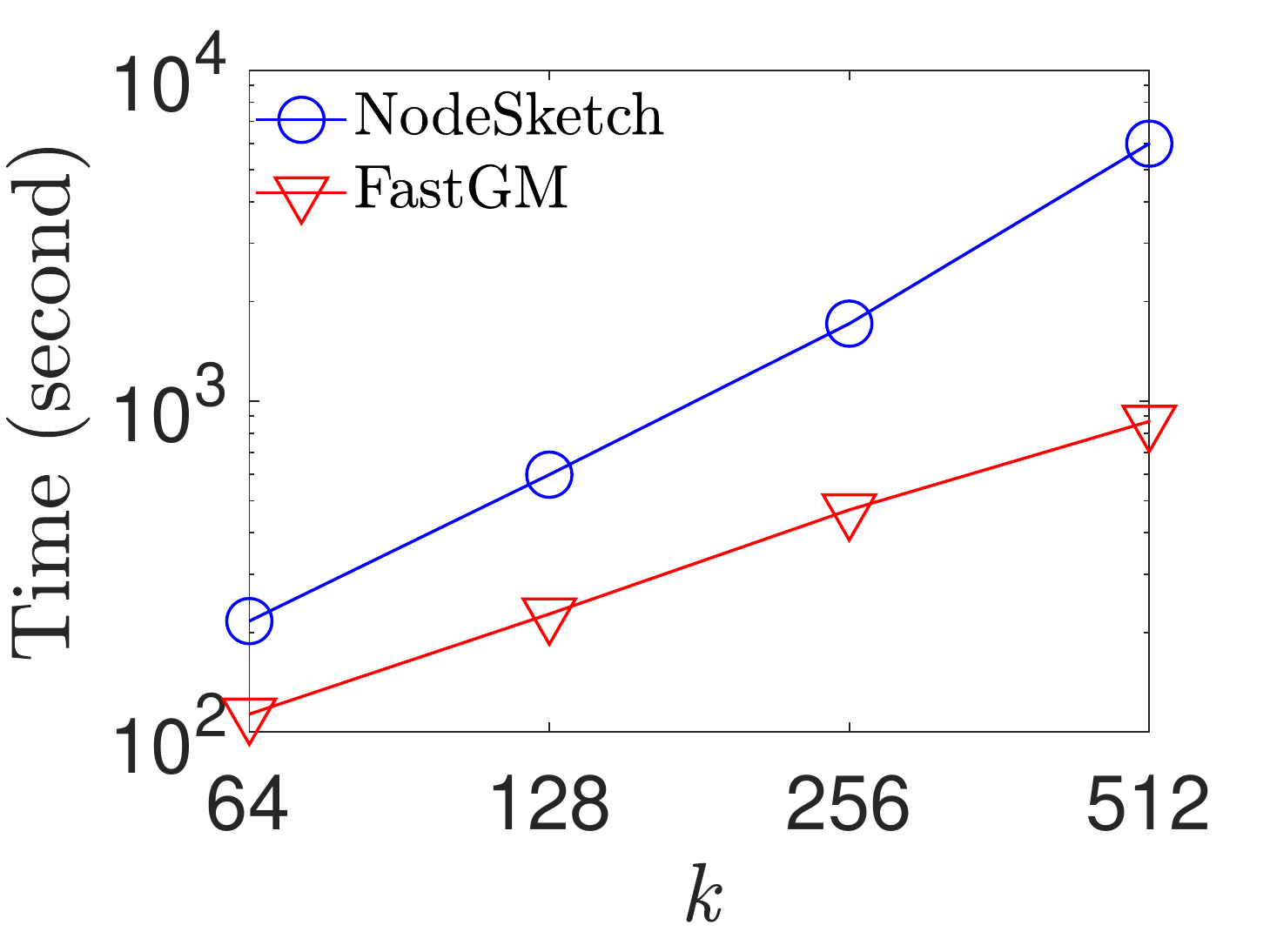}
		\caption{Twitter, total time}		
	\end{subfigure}
	\begin{subfigure}[t]{0.49\linewidth}
		\centering
		\includegraphics[width=\linewidth]{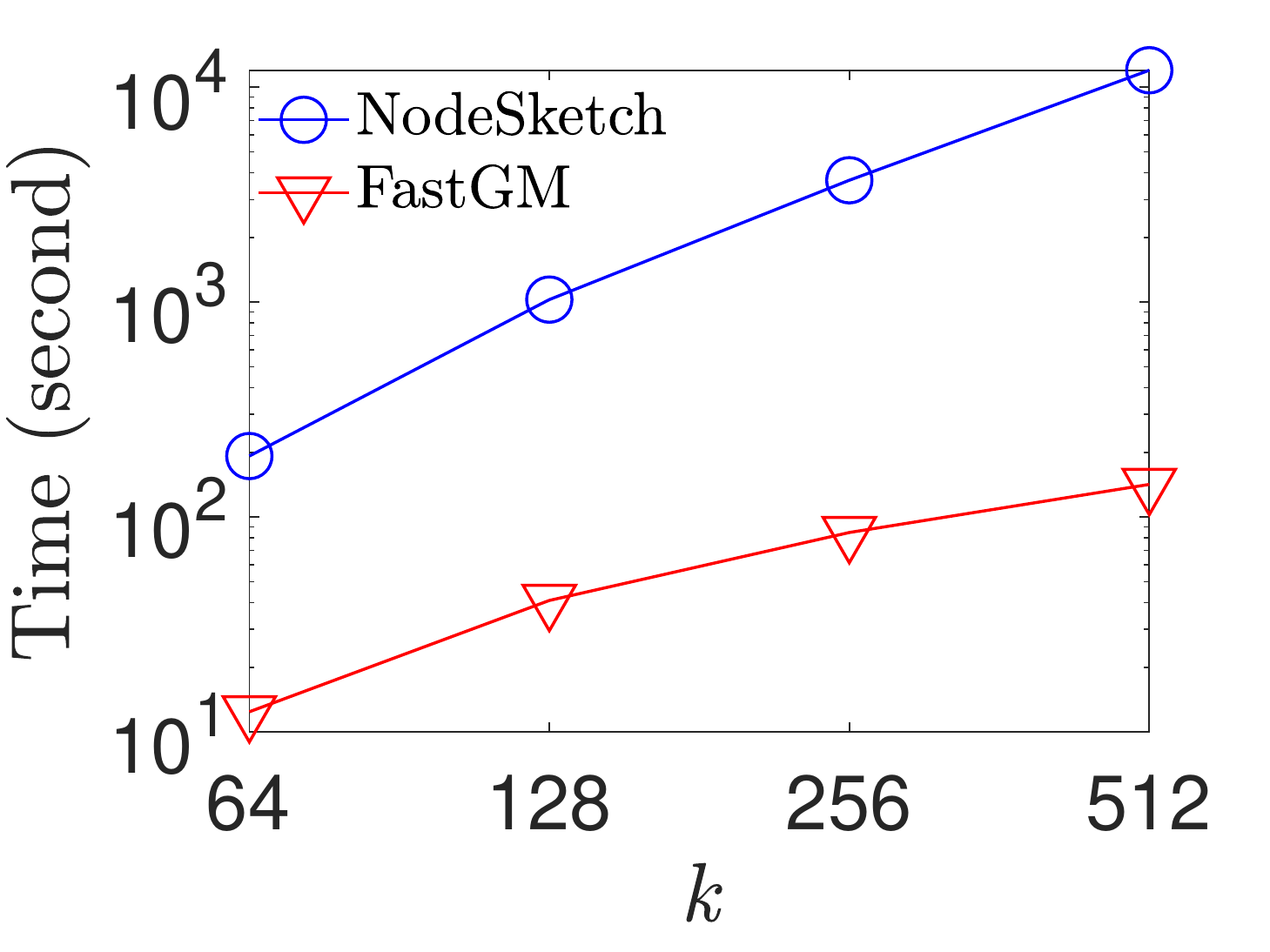}
		\caption{WikiTalk, sketching time}		
	\end{subfigure}
	\begin{subfigure}[t]{0.49\linewidth}
		\centering
		\includegraphics[width=\linewidth]{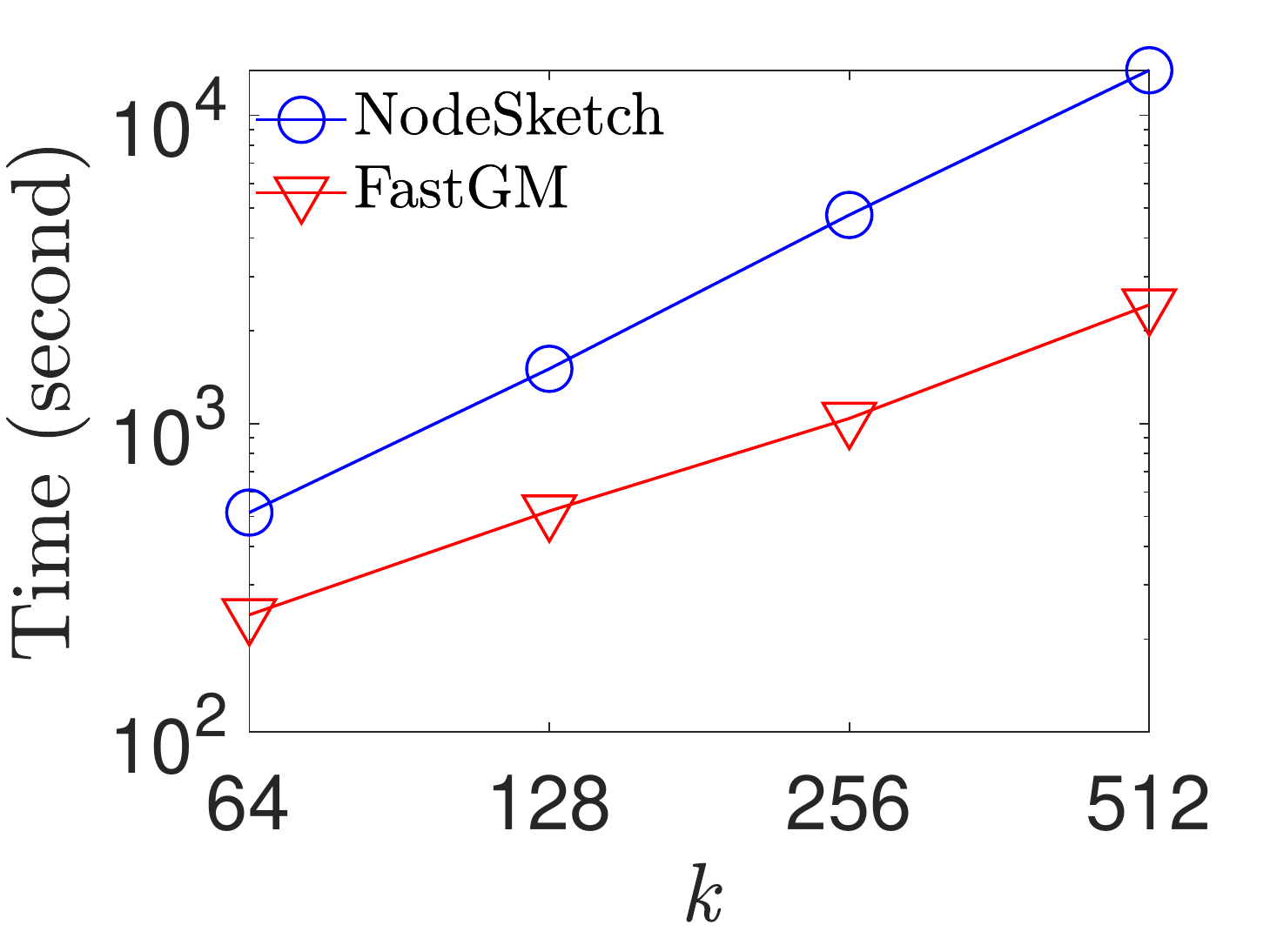}
		\caption{WikiTalk, total time}		
	\end{subfigure}
	\caption{(Task 2) Efficiency of our method FastGM in comparison with NodeSketch for different $k$.}\label{fig:real2}
\end{figure}

\subsection{Probability Jaccard Similarity Estimation}
We conduct experiments on both synthetic and real-world datasets for task 1. Specially, we first use synthetic weighted vectors to evaluate the performance of FastGM for vectors with different dimensions. Then, we show results on a variety of real-world datasets.

\textbf{Results on synthetic vectors.}
In this experiment, we also compare our method with another state-of-the-art algorithm BagMinHash \cite{ertl2018bagminhash}, which is used for estimating weighted Jaccard similarity.
We note that BagMinHash estimates an alternative similarity metric.
However, a lot of experiments and theoretical analysis \cite{moulton2018maximally} have shown that weighted Jaccard similarity and probability Jaccard similarity usually have similar performance on many applications such as fast searching similar set.
%Thus, we show that FastGM also outperforms BagMinHash on efficiency.
%The experiments are conducted on synthetic datasets where the weights in each vector follow uniform distribution and exponential distribution.
We conduct experiments on weighted vectors with uniform-distribution weights.
Without loss of generality, we let $n^+_{\vec{v}} = n$ for each vector, i.e., all elements of each vector are positive.
As shown in Figures~\ref{fig:syndataset} (a) and (b), when $n=10^3$, FastGM is $13$ and $22$ times faster than BagMinHash and $\mathcal{P}$-MinHash respectively.
As $n$ increases to $10^4$, the improvement becomes $8$ and $125$ times respectively.
Especially, the sketching time of our method is around $0.02$ seconds when $n=10^4$ and $k=2^{12}$, while BagMinHash and $\mathcal{P}$-MinHash take over $0.15$ and $2.5$ seconds for sketching respectively.
Figures~\ref{fig:syndataset} (c) and (d) show the running time of all competitors for different $n$.
Our method FastGM is 3 to 100 times faster than $\mathcal{P}$-MinHash for different $n$.
Compared with BagMinHash, FastGM is about 10 times faster when $n=1,000$,
and is comparable as $n$ increases to $100,000$.
It indicates that our method FastGM significantly outperforms BagMinHash
for vectors having less than  $100,000$ positive elements,
which are prevalent in real-world datasets.
%When $n$ is far larger than $k$ (i.e. Figure \ref{fig:syndataset} (b)), we can see that the sketching time of FastGM is almost irrelevant to the size of the sketch $k$.
%The reason is that the sketch is early filled within the first ``while'' iteration (Line 5 - 17) of LinearFill but our algorithm will not break until one ``while'' iteration finishes.
%Meanwhile, at the FastPrune module, most elements will stop generate balls at their first trials.
In addition, we also conduct experiments on weighted vectors with exponential-distribution weights and omit similar results here.

\textbf{Results on real-world datasets.}
Next, we show results on the real-world datasets in Table~\ref{tab:datasets}.
%In this experiment, we compare FastGM with $\mathcal{P}$-MinHash and BagMinHash on efficiency and with only $\mathcal{P}$-MinHash on effectiveness.
Figure \ref{fig:time_realdata} exhibits the sketching time of all algorithms.
We see that our method outperforms $\mathcal{P}$-MinHash and BagMinHash on all the datasets and the improvement increases as $k$ increases.
On sparse datasets such as Real-sim, Rcv1, and MovieLens, FastGM is about $8$ and $12$ faster than $\mathcal{P}$-MinHash and BagMinHash respectively.
BagMinHash is even slower than $\mathcal{P}$-MinHash on these datasets.
On datasets Webspam and Last.fm, we note that FastGM is $55$ and $80$ times faster than $\mathcal{P}$-MinHash respectively.
Figure \ref{fig:rmse} shows the estimation error of FastGM and $\mathcal{P}$-MinHash on datasets Real-sim and Webspam.
Due to the large number of vector pairs, we here randomly select $100,000$ pairs of vectors from each dataset and report the average RMSE.
We note that both algorithms give similar accuracy, which is coincident with our analysis.
We omit similar results on other datasets.
\begin{figure}[t]
	\centering
	\begin{subfigure}[t]{0.49\linewidth}
		\centering
		\includegraphics[width=\linewidth]{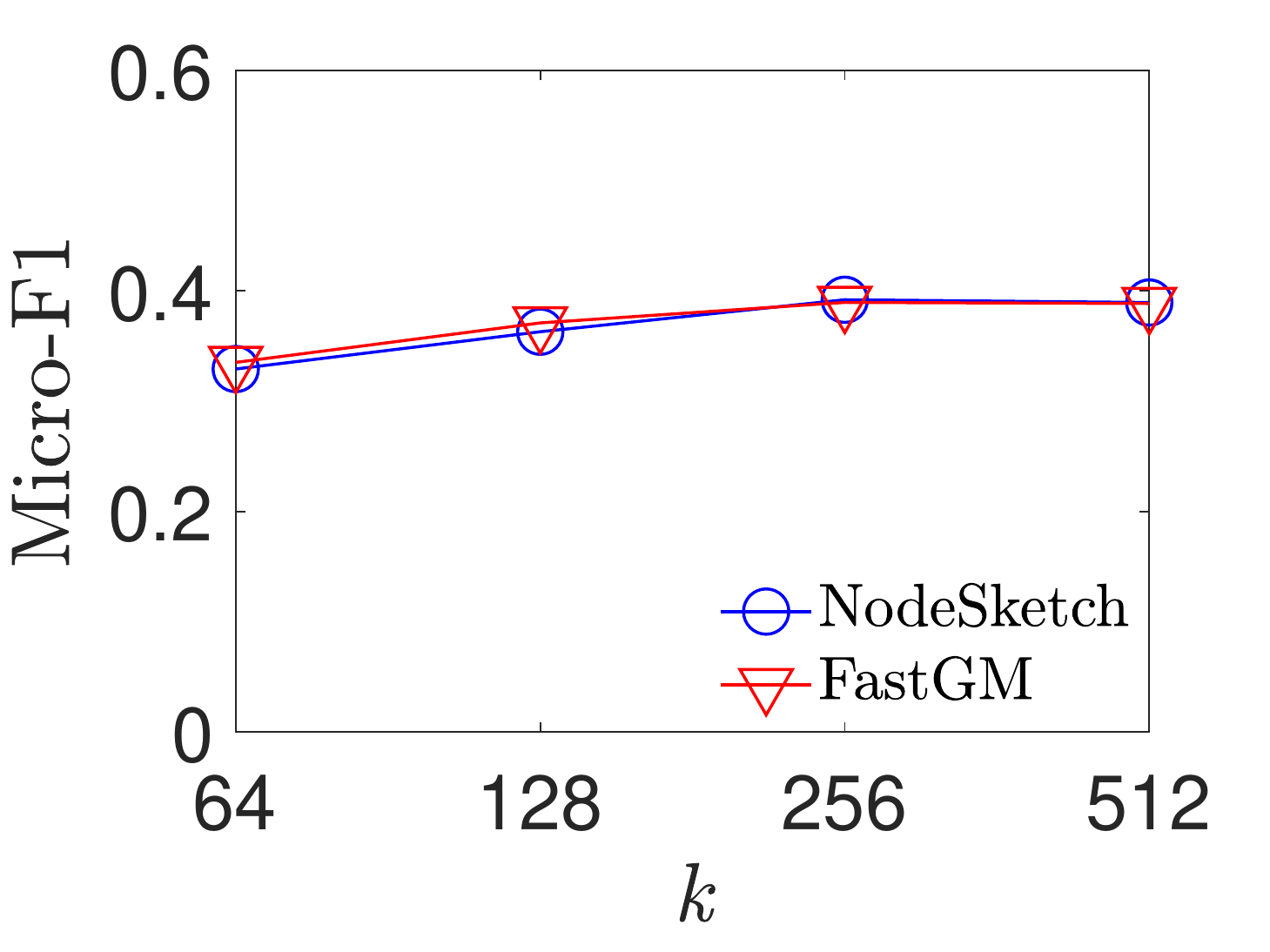}
		\caption{Micro-F1}		
	\end{subfigure}
	\begin{subfigure}[t]{0.49\linewidth}
		\centering
		\includegraphics[width=\linewidth]{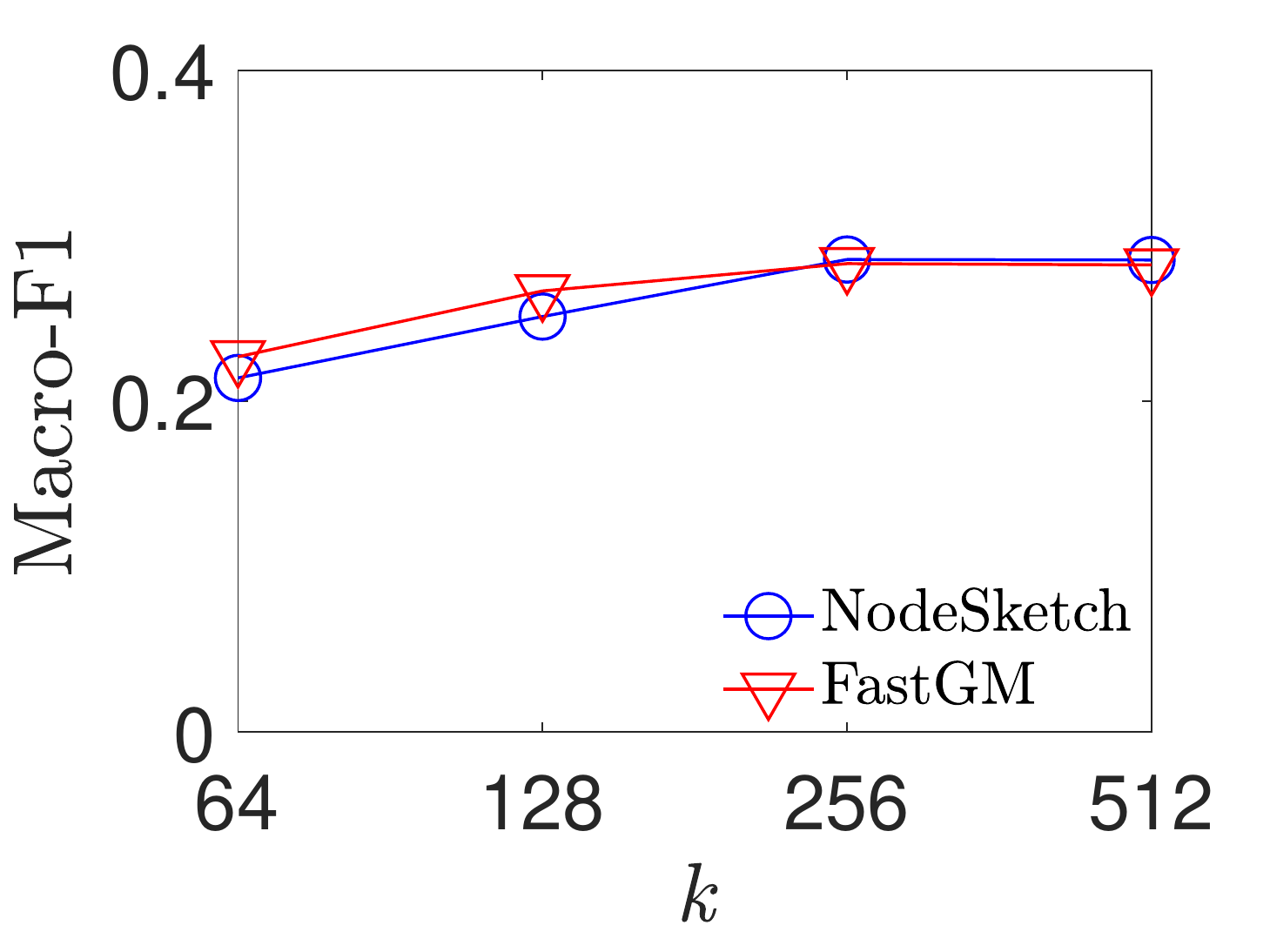}
		\caption{Macro-F1}
	\end{subfigure}
	\begin{subfigure}[t]{0.49\linewidth}
		\centering
		\includegraphics[width=\linewidth]{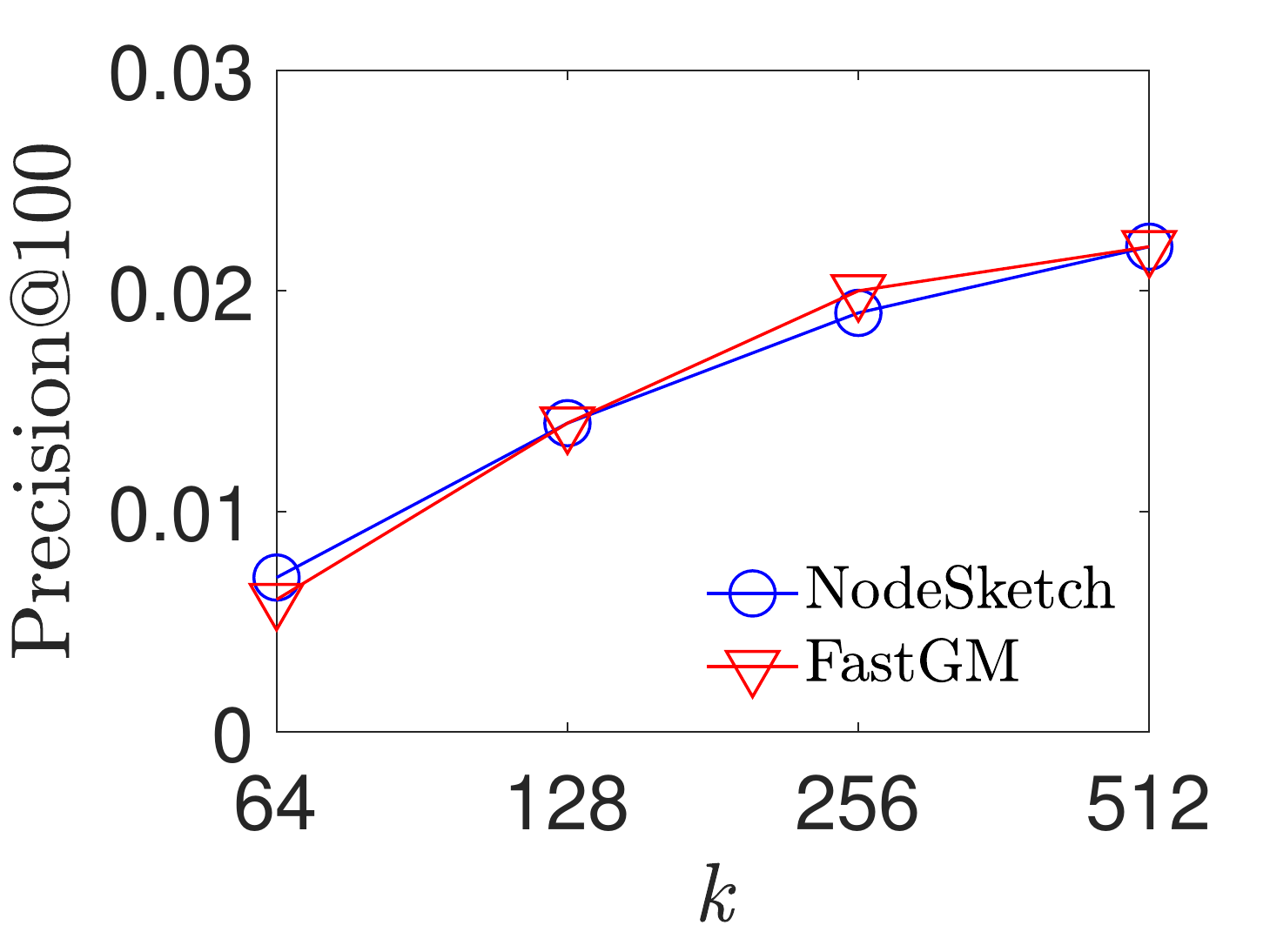}
		\caption{Precision$@100$}	
	\end{subfigure}
	\begin{subfigure}[t]{0.49\linewidth}
		\centering
		\includegraphics[width=\linewidth]{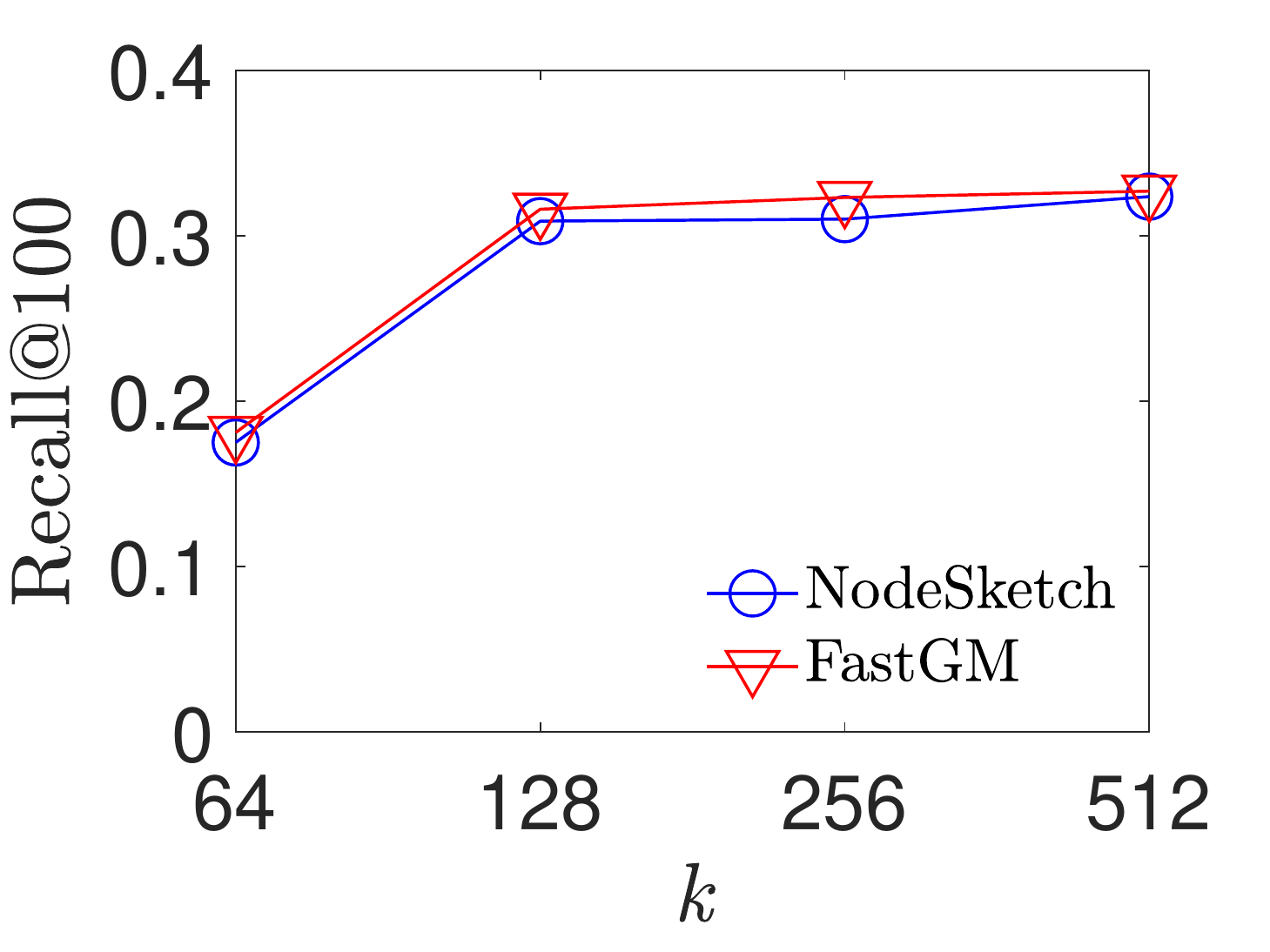}
		\caption{Recall$@100$}
	\end{subfigure}
	\caption{(Task 2, YouTube) Accuracy of FastGM and NodeSketch. (a)(b): node classification; (c)(d): link prediction.}\label{fig:classify}
\end{figure}

\subsection{Graph Embedding}
We compare FastGM with the regular NodeSketch to demonstrate the efficiency of our method on the graph embedding task.
Specially, we show the sketching time and the total time for different $k$ (i.e., the size of node embeddings), where the sketching time refers to the accumulated time of computing Gumbel-Max sketches from different orders of the self-loop-augmented adjacent matrix.
As shown in Figure \ref{fig:real2},  we note that our method FastGM gives significant improvement on all datasets.
In detail, on dataset WikiTalk, FastGM gives an improvement of $16$ times at $k=2^6$ and the improvement increases to $84$ times at $k=2^{9}$ on sketching time,
which results in a gain of up to $5$ times improvements for the total time of learning node embeddings.
%We also notice that both algorithms give similar esults on merging time because both algorithms share the same merging procedure.

We also conduct experiments on two popular applications of graph embedding, i.e., node classification and link prediction.
All experimental settings are the same as \cite{yang2019nodesketch}.
For node classification, we randomly select 10\% nodes as the training set and others as the testing set.
Then, we build a one-vs-rest SVM classifier based on the training set.
For link prediction, we randomly drop out 20\% edges from the original graph as the testing set and learn the embeddings based on the remaining graph.
We predict the potential edges by generating a ranked list of node pairs.
For each node pair, we use the Hamming similarity of their embeddings to generate the ranked list.
Due to the massive size of node pairs, we randomly sample $10^5$ pairs of nodes for evaluation.
We report Precision$@100$ and Recall$@100$.
Figure \ref{fig:classify} shows the results on node classification as well as link prediction on dataset YouTube.
We notice that our method FastGM gives similar accuracy compared with NodeSketch.
We omit the results of the other three datasets.

\section{Conclusions and Future Work} \label{sec:conclusions}
In this paper, we develop a novel algorithm FastGM to fast compute
a nonnegative vector's Gumbel-Max sketch, which consists of $k$ independent Gumbel-Max variables.
We prove that FastGM generates Gumbel-Max sketch with the same quality as the traditional Gumbel-Max Trick but reduces time complexity from $O(n^+ k)$ to $O(k \ln k + n^+)$,
where $n^+$ is the number of the vector's positive elements.
We conduct a variety of experiments on two tasks: Probability Jaccard similarity estimation and graph embedding,
and the experimental results demonstrate that our method FastGM is orders of magnitude faster than the state-of-the-arts without sacrificing accuracy.
In the future, we plan to extend FastGM to vectors consisting of elements arriving in a streaming fashion.

\section*{Acknowledgment}
The research presented in this paper is supported in part by  National Key R\&D Program of China (2018YFC0830500), Shenzhen Basic Research Grant (JCYJ20170816100819428), National Natural Science Foundation of China (61922067, U1736205, 61902305),  MoE-CMCC ``Artifical Intelligence'' Project (MCM20190701), Natural Science Basic Research Plan in Shaanxi Province of China (2019JM-159), Natural Science Basic Research Plan in ZheJiang Province of China (LGG18F020016).

\balance
\bibliographystyle{unsrt}
\bibliography{COPH,randpe,ctstream,simcar,albitmap,dynamic}

\end{document}